\documentclass{article}

\usepackage{arxiv}

\usepackage[utf8]{inputenc} 
\usepackage[T1]{fontenc}    
\usepackage{hyperref}       
\usepackage{url}            
\usepackage{booktabs}       
\usepackage{amsfonts}       
\usepackage{nicefrac}       
\usepackage{microtype}      
\usepackage{lipsum}		
\usepackage{graphicx}
\usepackage{natbib}
\usepackage{doi}
\usepackage{colortbl}

\title{Bayesian outcome weighted learning}

\author{ \href{https://orcid.org/0000-0000-0000-0000}{\includegraphics[scale=0.06]{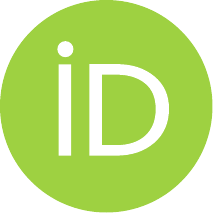}\hspace{1mm}Sophia Yazzourh}\\
	Institut de Mathématiques de Toulouse\\
	UMR5219 - Université de Toulouse\\
	CNRS - UPS IMT \\
    F-31062 Toulouse Cedex 9, France\\
	\texttt{sophia.yazzourh@math.univ-toulouse.fr} \\
	\And
	\href{https://orcid.org/0000-0000-0000-0000}{\includegraphics[scale=0.06]{orcid.pdf}\hspace{1mm}Nikki L. B. Freeman} \\
	Department of Biostatistics and Bioinformatics\\
    Duke Clinical Research Institute\\
	Duke University\\
	Durham, North Carolina 27701 USA \\
	\texttt{nikki.freeman@duke.edu} \\
}

\date{}

\renewcommand{\shorttitle}{Bayesian Outcome Weighted Learning}

\hypersetup{
pdftitle={Bayesian outcome weighted learning},
pdfsubject={},
pdfauthor={David S.~Hippocampus, Elias D.~Striatum},
pdfkeywords={Precision medicine, Individualized treatment rule, Bayesian machine learning},
}

\usepackage{ulem}
\usepackage{amsmath}
\usepackage[capitalise, noabbrev]{cleveref}
\usepackage{amssymb}
\usepackage{xcolor}
\usepackage{bbm}
\usepackage{tcolorbox}
\usepackage{bbold}

\newcommand{\matrixSymbol}[1]{\mathbf{#1}} 
\newcommand{\dtr}{d}
\newcommand{\dtrOpt}{d^{\text{opt}}}
\newcommand{\valuefnx}{V}
\newcommand{\betaVector}{\boldsymbol{\beta}}
\newcommand{\lambdaVector}{\boldsymbol{\lambda}}
\newcommand{\omegaVector}{\boldsymbol{\omega}}
\newcommand{\gammaVector}{\boldsymbol{\gamma}}
\newcommand{\reward}{R}
\newcommand{\rewardObserved}{r}
\newcommand{\rewardVectorObserved}{\mathbf{r}}
\newcommand{\action}{A}
\newcommand{\actionObserved}{a}
\newcommand{\actionVectorObserved}{\mathbf{a}}
\newcommand{\newAction}{\tilde{a}}
\newcommand{\propensityScore}{\rho}
\newcommand{\fullCovariateVector}{X}
\newcommand{\fullCovariateVectorObserved}{\mathbf{x}}
\newcommand{\fullCovariateMatrix}{\matrixSymbol{X}}
\newcommand{\newCovariate}{\tilde{\mathbf{x}}}

\newcommand{\ssParamVector}{\boldsymbol{\gamma}}
\newcommand{\betaPriorMean}{\boldsymbol{\mu}_0}
\newcommand{\betaPriorVar}{\sigma^2_0}

\newtcolorbox{mybox}{
    colback=white,  
    colframe=black, 
    sharp corners, 
    boxrule=0.5pt, 
    before skip=10pt, 
    after skip=10pt, 
    left=10pt, 
    right=10pt 
}

\definecolor{lightblue}{rgb}{0.8,0.85,1}

\begin{document}
\maketitle

\begin{abstract}
	One of the primary goals of statistical precision medicine is to learn optimal individualized treatment rules (ITRs). The classification-based, or machine learning-based, approach to estimating optimal ITRs was first introduced in outcome-weighted learning (OWL).  OWL recasts the optimal ITR learning problem into a weighted classification problem, which can be solved using machine learning methods, e.g., support vector machines. In this paper, we introduce a Bayesian formulation of OWL. Starting from the OWL objective function, we generate a pseudo-likelihood which can be expressed as a scale mixture of normal distributions. A Gibbs sampling algorithm is developed to sample the posterior distribution of the parameters. In addition to providing a strategy for learning an optimal ITR, Bayesian OWL provides a natural, probabilistic approach to estimate uncertainty in ITR treatment recommendations themselves. We demonstrate the performance of our method through several simulation studies. 
\end{abstract}


\section{Introduction}
The task of statistical precision medicine is to learn from data how to match patients to treatments with the aim of improving health outcomes \citep{Kosorok2019-ib}. One way to operationalize this goal is through individualized treatment regimes (ITRs), functions that map from patient characteristics to treatment recommendations. Ideally, we would like to learn ITRs that if followed in practice would lead to better outcomes on average in the target population than if another treatment strategy was used, e.g., a one-size-fits-all approach. These ITRs are called optimal ITRs \citep{Kosorok2019-ib, Murphy2003-xo}.

In the language of reinforcement learning, we focus on the "batch, off-policy" setting. By ``batch" we mean that data have been previously collected and no new data will be received, and by ``off policy' we mean that the strategy for assigning treatments in the observed data (e.g., through randomization as in a clinical trial) may not be the optimal strategy (or alternatively, regime or policy) \citep{Sutton2018-ij}. Within this setting, a large number of methods and approaches have been developed to learn such ITRs from data. 

Some approaches estimate the expected value of the outcome we would expect under a particular ITR without any parametric assumptions. Then, the optimal ITR may be learned by searching over a class of ITRs. Examples of this general strategy include those proposed by \cite{Orellana2010-zu}, \cite{Robins2008-mi}, and \cite{Zhang2013-qp}.

Another class of approaches, sometimes referred to as indirect methods, model the mean of the outcome conditional on treatment and covariates. For a multi-stage ITR, sometimes called a dynamic treatment regime, this entails specifying a conditional mean model for each stage. Through these estimated conditional means, optimal ITRs can be deduced. One of the most popular regression-based frameworks for learning optimal ITRs is Q-learning \citep{Watkins1989-er, Murphy2005-su}. Q-learning has been used to learn optimal ITRs in many settings, including clinical trial data \citep{Schulte2014-vp}, observational data \citep{Moodie2012-jh}, and in the presence of censoring for time-to-event data \cite{Goldberg2012-mk}. 

Machine learning or classification-based optimal ITR learning approaches convert the optimal ITR learning problem into the classification framework by which machine learning methods can be employed. \cite{Zhao2012-vc} introduced outcome weighted learning (OWL) which leverages a simple value function estimator and the Radon-Nikodym theorem to rewrite the value function as a weighted classification problem. Consequently, learning the ITR that optimizes the value function can be solved as minimizing the classification loss function. Since its introduction, a number of extensions to OWL have been made including backwards outcome weighted learning (BOWL) and simultaneous outcome weighted learning (SOWL) for learning optimal multi-stage treatment regimes \citep{Zhao2015-gb}, residual weighted learning (RWL) \citep{Zhou2017-go} and augmented outcome-weighted learning (AOL) \citep{Liu2018-so} which improve the finite sample properties of OWL, robust outcome weighted learning (ROWL) which uses an angle-based classification approach \citep{Fu2019-iz}, and efficient augmentation and relaxation learning (EARL) which employs both a propensity model and outcome model and has the double robustness property \cite{Zhao2019-ky}.

Finally, a few Bayesian approaches for learning optimal ITRs have also been proposed. The Bayesian machine learning (BML) approach was introduced by \cite{Murray2018-bx}. It employs Bayesian modeling within a framework that closely aligns with Q-learning by modeling the outcomes at each stage. Likelihood-based approaches, strategies that models both the distribution of the final outcome and the intermediate outcomes, have also been proposed within the Bayesian framework \citep{Thall2002-hg, Thall2007-gr, Arjas2010-dp, Zajonc2012-gz, Xu2016-qm, Yu2023-fg}. 

The focus of this paper will be on the machine learning, or classification-based method, for learning optimal ITRs. Although classification-based approaches are powerful and avoid estimating models that are not the target of the analysis itself, there are limitations. For example, many machine learning methods for classification do not naturally quantify uncertainty, e.g., quantification of the uncertainty of a particular prediction. While this may be acceptable in some cases, being unable to quantify uncertainty is a serious gap when generating evidence for health care decision-making. Moreover, machine learning analyses are often evaluated in terms of predictive power which does not necessarily translate into inferential capability. 

In this paper, we present a Bayesian approach to OWL. To our knowledge, this is the first Bayesian optimal ITR learning strategy to directly learn optimal ITRs. Using a construction similar to \cite{Polson2011-rw}, we construct a pseudo-likelihood from the weighted classification loss function. Once transformed from an optimization-based framework to a probabilistic framework, our method generates an entire posterior distribution that can be used for inference and, most powerfully, for uncertainty quantification of the treatment recommendations themselves. Our main contributions are as follows:
\begin{enumerate}
    \item We propose a Bayesian approach to learning optimal ITRs that leverages the classification-based framework and avoids modeling the outcome or nuisance conditional mean models.
    \item We propose a simple Gibbs sampling algorithm for learning such an optimal ITR.
    \item We demonstrate how to use our resulting pseudo-posterior distribution to quantify uncertainty in the treatment recommendations. 
\end{enumerate}
In \cref{sec:setting}, we set the notation and review OWL and Bayesian support vector machines. In \cref{sec:approach} we construct the probabilistic formulation of the OWL classification problem, derive a Gibbs sampling algorithm for estimation, and detail our approach to uncertainty quantification. In \cref{sec:simulation} we demonstrate the performance of our approach through simulation studies. 
We conclude in \cref{sec:discussion} with a discussion of our results and future work.

\section{Background}\label{sec:setting}
\subsection{Setting}
We let $\action \in \mathcal{A}=\{-1, 1\}$ denote the action, or treatment, and assume that observed treatments are assigned randomly as in a clinical trial with $P(A = 1) = \propensityScore$ known. Let $\fullCovariateVector_i = (X_{i,1}, \ldots, X_{i,p})^\top \in \mathcal{X}$ denote the p-dimensional biomarker and prognostic information vector, and let $\reward$ denote the outcome (bigger is better). We further assume that the reward can be rescaled so that $\reward >0$. Then, the observed data is iid replicates of $(\action_i, \fullCovariateVector_i, \reward_i)$ for $i = 1, \ldots, n$.

An ITR is a function $\dtr$ that maps from $\mathcal{X}$ to a recommended treatment in $\mathcal{A}$. For a given ITR $d$, the value of $\dtr$ is $\valuefnx(d) = \mathbb{E}[\reward(\dtr)]$, where $\reward(\dtr)$ is the reward we would observe if treatments were allocated according to rule $\dtr$. An optimal ITR $\dtrOpt$ satisfies $\valuefnx(\dtrOpt) \ge \valuefnx(\dtr)$ for all $\dtr \in \mathcal{D}$, where $\mathcal{D}$ is a class of ITRs. Our goal is to learn an optimal ITR $\dtrOpt$. Under the assumptions of causal consistency, the stable unit treatment value assumption, no unmeasured confounding, and positivity, $\valuefnx(d)$ can be identified from the observed data and $\valuefnx(d) = \mathbb{E}\{\max_{\action \in \mathcal{A}} \mathbb{E}[\reward \vert \action = \dtr(\fullCovariateVectorObserved), \fullCovariateVector = \fullCovariateVectorObserved]\}$. 

\subsection{Outcome weighted learning}
If we let $P$ denote the distribution of $(\fullCovariateVector, \action, \reward)$, and $P^\dtr$ denote the distribution of $(\fullCovariateVector, \action, \reward)$ when $\action = \dtr(\fullCovariateVector)$, then the reward we would expect if ITR $\dtr(\fullCovariateVector)$ were followed is given by
\begin{align}\label{eq:change_of_measure}
    \mathbb{E}^\dtr(\reward) = \int \reward dP^\dtr = \int R \frac{dP^\dtr}{dP} dP = \mathbb{E}\left[\frac{\mathbbm{1}(\action = \dtr(\fullCovariateVector))}{\action \propensityScore + (1-\action)/2} \reward \right].
\end{align}
\cite{Zhao2012-vc} showed that maximizing \cref{eq:change_of_measure} is equivalent to a weighted classification problem and thereby solvable using techniques from machine learning. Specifically, they proposed OWL, a strategy for learning an optimal ITR using a convex surrogate loss function in place of the zero-one loss function and strategies from support vector machines. OWL minimizes the objective function
\begin{align}\label{eq:OWL_objective_fnx}
    Q_n^{\text{OWL}}(\boldsymbol{\beta}) = \frac{1}{n} \sum_{i = 1}^n \frac{\reward_i}{\action_i \propensityScore + (1-\action_i)/2}(1-\action_i h(\fullCovariateVector_i, \boldsymbol{\beta}))_+
\end{align}
where $(z)_+ = max(z, 0)$ denotes the hinge loss function and $h(\cdot)$ is the ITR parameterized by $\boldsymbol{\beta}$. \cite{Song2015-yk} introduced a penalized variant of OWL that included a regularization term $p_{\lambda}(\betaVector)$ for the ITR parameters. POWL minimizes the objective function
\begin{align}\label{eq:POWL_objective_fnx}
  Q_n^{\text{POWL}}(\betaVector) \frac{1}{n} \sum_{i = 1}^n \frac{\reward_i}{\action_i \propensityScore + (1-\action_i)/2}(1-\action_i h(\fullCovariateVector_i, \boldsymbol{\beta}))_+ + \sum_{j=1}^{p} p_{\lambda}(\vert \beta_j \vert) 
\end{align}
where $p_\lambda(\betaVector)$ is a penalty function and $\lambda$ is a tuning parameter. 

\subsection{Bayesian support vector machines}
Although the pure machine learning framework is powerful, it is limited in its ability to capture and model uncertainty as in a statistical framework. \cite{Polson2011-rw} bridged this gap between pure machine learning and statistical modeling for SVMs by showing how to cast SVM into a Bayesian framework. They considered the $L^{\alpha}$-norm regularized support vector classifier that chooses $\beta$ to minimize
\begin{align}\label{eq:linear_svm_objective_fnx}
    d_\alpha(\beta, \nu) = \sum_{i = 1}^n \max (1 - r_i \mathbf{x}_i^\top \beta, 0) + \nu^{-\alpha} \sum_{j = 1}^{k}\vert \beta_j/\sigma_j \vert^{\alpha}
\end{align}
where $\sigma_j$ is the standard deviation of the $j$-th element of $\mathbf{x}$ and $\nu$ is a tuning parameter. For this objective function, the learned classifier is a linear classifier. \cite{Polson2011-rw} shows that minimizing \cref{eq:linear_svm_objective_fnx} is equivalent to finding the mode of the pseudo-posterior distribution $p(\beta \vert \nu, \alpha, y)$
\begin{align}
    p(\beta \vert \nu, \alpha, r) &\propto \exp(-d_{\alpha}(\beta, \nu)) \nonumber \\
    & \propto C_{\alpha}(\nu) L(r \vert \beta) p(\beta \vert \nu, \alpha)
\end{align}
where $C_{\alpha}$ is a pseudo-posterior normalization constant. Thus, the data dependent factor $L(y \vert \beta)$ is a pseudo-likelihood
\begin{align}
    L(r \vert \beta) = \prod_{i} L_{i}(r_{i} \vert \beta) = \exp \left\{ x-2 \sum_{i = 1}^{n} \max(1 - r_i x_i^\top \beta, 0) \right\}.
\end{align}
The main theoretical result from \Citet{Polson2011-rw} is that the pseudo-likelihood contribution $L_i(r_i \vert \beta)$ is a location-scale mixture of normals (\cite{Polson2011-rw}, Theorem 1). 

\section{Our approach}\label{sec:approach}
We follow the strategy employed by \Citet{Polson2011-rw} to cast the OWL objective function into a probabilistic Bayesian learning framework. The conversion is not one-to-one since \Citet{Polson2011-rw} constructed a Bayesian model for a standard SVM whereas the objective function for OWL \cref{eq:OWL_objective_fnx} is a weighted SVM problem. We first employ a non-penalty prior to mimic original formulation of OWL as in \cite{Zhao2012-vc}, and later we demonstrate the inclusion of a penalty prior on the the ITR coefficients. Minimizing \cref{eq:OWL_objective_fnx} is equivalent to finding the mode of the pseudo-posterior which we can write as
\begin{align}
    p(\fullCovariateVectorObserved \vert a_i, \nu, \alpha) & \propto \exp(-Q_n(\boldsymbol{\beta}, \nu, \alpha)) \nonumber \\
    & \propto \exp \left\{\sum_{n = 1}^n \frac{\rewardObserved_i}{\actionObserved_i \propensityScore + (1 - \action_i)/2} (1 - \actionObserved_i h(\fullCovariateVectorObserved_i, \betaVector))_{+}\right\} \prod_{j=1}^p p(\beta_j \vert \mu_0, \sigma^2_0)  \nonumber \\
    & \propto C(\nu, \alpha) L(a \vert \betaVector) p(\betaVector \vert \mu_0, \sigma^2_0).
\end{align}
Throughout, we will assume $\reward >0$. When this is not the case, a distance-preserving transformation of $\reward$ from $\mathbb{R}$ to $\mathbb{R}^+$ can be used. Assuming that $h$ is linear, i.e., $h(\fullCovariateVectorObserved_i, \betaVector) = \fullCovariateVectorObserved_i^\top \betaVector$ and following the strategy taken in Theorem 1 of \Citet{Polson2011-rw}, the contribution of a single observation to the pseudo-likelihood is given by

\begin{align}\label{eq:powl_pseudolikelihood_contrib}
    L_i(\actionObserved_i \vert \rewardObserved_i, \fullCovariateVectorObserved_i, \betaVector) =& \exp\left\{ -2 \frac{\rewardObserved_i}{\actionObserved_i \propensityScore + (1 - \actionObserved_i)/2} \max(1 - \actionObserved_i \fullCovariateVectorObserved_i^\top \betaVector, 0)\right\} \nonumber \\
    =& \mathbbm{1}(\actionObserved_i = 1) \int_{0}^{\infty} \frac{1}{ \sqrt{ 2 \pi \lambda_i}} \exp \left\{-\frac{1}{2 \lambda_i} \left(\frac{\rewardObserved_i}{\propensityScore} + \lambda_i - \frac{\rewardObserved_i}{\rho} \actionObserved_i \fullCovariateVectorObserved_i^\top \betaVector \right)^2 \right\} d\lambda_i  \nonumber\\
    &+ \mathbbm{1}(\actionObserved_i = -1) \int_{0}^{\infty} \frac{1}{\sqrt{2 \pi \lambda_i}} \exp \left\{-\frac{1}{2 \lambda_i} \left(\frac{\rewardObserved_i}{1-\propensityScore} +\lambda_i - \frac{\rewardObserved_i}{1-\propensityScore} \actionObserved_i \fullCovariateVectorObserved_i^\top \betaVector \right)^2 \right\} d\lambda_i,    
\end{align}
or in other words that $L_i(\actionObserved_i, \lambda_i \vert \rewardObserved_i, \fullCovariateVectorObserved_i, \betaVector)$ is a scale mixture of Gaussians (full details are contained in \cref{appendix1}).

\subsection{Prior specification for the ITR parameters}
In their formulation of Bayesian SVM, \cite{Polson2011-rw} use the exponential power prior for $\betaVector$, a prior that can be show to be equivalent to L1-regularization of the regression parameters. Regularization of the OWL parameters have been explored as in \cite{Song2015-yk}. In this paper, we first construct our method as an analogy to the original formulation of OWL without penalization. We make this choice because (1) our primary aim is to develop a Bayesian classification-based ITR learning approach, and because (2) L1-regularization does not necessarily yield sparse rules (see the discussion in Section 4.1 of \cite{Polson2011-rw}). However, regularization help avoid overfitting, a common problem in machine learning. Thus, we also explore penalty priors for $\betaVector$, including the exponential power prior distribution and the spike-and-slab prior distribution.

\subsubsection{Normal prior distribution for \texorpdfstring{$\betaVector$}{betaVector}}

We first consider the case with normal distribution priors on the treatment rule parameters $\beta_j \sim N(\mu_0, \sigma^2_0)$ for $j = 1, \ldots, p$ where $\mu_0$ and $\sigma^2_0$ are hyperparameters. With the pseudo-likelihood and a suitable prior for the ITR parameters defined, we can write the pseudo-posterior distribution as
\begin{align}\label{eq:full_pseudoposterior_normalprior}
    p(\betaVector, \lambdaVector \vert \fullCovariateVectorObserved, \rewardVectorObserved, \actionVectorObserved, \alpha, \nu) \propto& \prod_{\{i: \actionObserved_i = 1 \}}^n \lambda_i^{-1/2} \cdot  \exp\left\{-\frac{1}{2} \sum_{i: \actionObserved_1 = 1}^n \frac{\left(\frac{\rewardObserved_i}{\propensityScore} + \lambda_i - \frac{\rewardObserved_i}{\rho} \actionObserved_i \fullCovariateVectorObserved_i^\top \betaVector \right)^2}{\lambda_i}  \right\}  \nonumber \\
    &\times \prod_{\{i: \actionObserved_i = -1 \}}^n \lambda_i^{-1/2} \cdot  \exp \left\{-\frac{1}{2} \sum_{i: \actionObserved_i = -1}^n  \frac{\left(\frac{\rewardObserved_i}{1-\propensityScore} +\lambda_i - \frac{\rewardObserved_i}{1-\propensityScore} \actionObserved_i \fullCovariateVectorObserved_i^\top \betaVector \right)^2}{\lambda_i}  \right\} \nonumber \\
    &\times \prod_{j = 1}^p \frac{1}{\sqrt{2 \pi \sigma^2_0}} \exp\left\{-\frac{1}{2} \frac{(\beta_j - \mu_{0, j})^2}{\sigma^2_0}\right\}
\end{align}
where $\lambdaVector = (\lambda_1, \ldots, \lambda_n)^\top$, $\rewardVectorObserved = (\rewardObserved_i, \ldots, \rewardObserved_n)^\top$, and $\actionVectorObserved = (\actionObserved_1, \ldots, \actionObserved_n)^\top$.

\subsection{Exponential power prior distribution for \texorpdfstring{$\betaVector$}{betaVector}}
Rather than use normal priors for the coefficients of the rule, \cite{Polson2011-rw} employed an exponential power prior on $\betaVector$. This prior contains the regularization penalty, and from Theorem 2 of \cite{Polson2011-rw},
the double exponential prior regularization penalty can be written as 
\begin{align}
    p(\beta_j \vert \nu, \alpha = 1) =& \int_{0}^{\infty} \phi(\beta_j \vert 0, \nu^2 \omega_j \sigma_j^2) \frac{1}{2} e^{-\frac{\omega_j}{2}} d\omega_j 
\end{align}
where $p(\omega_j \vert \alpha) \propto \omega_j^{-\frac{3}{2}}St^+_{\alpha/2}(\omega_j^{-1})$ and $St^+_{\alpha/2}$ is the density function of a positive stable random variable of index $\alpha/2$. In particular, when $\alpha = 1$, $p(\omega_j \vert \alpha) \sim Exponential(2)$ (Corollary 1 of \Citet{Polson2011-rw}).

Under this prior distribution specification, we can write the pseudo-posterior distribution as
\begin{align}\label{eq:full_pseudoposterior_exponentialPowerPrior}
    p(\betaVector, \lambdaVector, \omegaVector \vert \fullCovariateVectorObserved, \rewardVectorObserved, \actionVectorObserved, \alpha, \nu) \propto& \prod_{\{i: \actionObserved_i = 1 \}}^n \lambda_i^{-1/2} \cdot  \exp\left\{-\frac{1}{2} \sum_{i: \actionObserved_1 = 1}^n \frac{\left(\frac{\rewardObserved_i}{\propensityScore} + \lambda_i - \frac{\rewardObserved_i}{\rho} \actionObserved_i \fullCovariateVectorObserved_i^\top \betaVector \right)^2}{\lambda_i}  \right\}  \nonumber \\
    &\times \prod_{\{i: \actionObserved_i = -1 \}}^n \lambda_i^{-1/2} \cdot  \exp \left\{-\frac{1}{2} \sum_{i: \actionObserved_i = -1}^n  \frac{\left(\frac{\rewardObserved_i}{1-\propensityScore} +\lambda_i - \frac{\rewardObserved_i}{1-\propensityScore} \actionObserved_i \fullCovariateVectorObserved_i^\top \betaVector \right)^2}{\lambda_i}  \right\} \nonumber \\
    &\times \prod_{j = 1}^p \omega_j^{-\frac{1}{2}} \cdot  \exp \left\{ -\frac{1}{2 \nu^2} \sum_{j = 1}^p \frac{\beta_j^2}{\sigma^2_j \omega_j}\right\} \cdot \prod_{j = 1}^p p(\omega_j \vert \alpha).
\end{align}
where $\omegaVector = (\omega_1, \ldots, \omega_p)^\top$.

\subsection{Spike-and-slab prior distrbituion for \texorpdfstring{$\betaVector$}{betaVector}}\label{eq:full_pseudoposterior_SSprior}
\cite{Polson2011-rw} also explored the use of a spike-and-slab prior for $\betaVector$. The spike-and-slab prior is a Bayesian approach used for variable selection. It combines a "spike" component, which is a Dirac delta function at zero, to induce sparsity by shrinking some coefficients exactly to zero, and a "slab" component that allows other coefficients to vary freely \citep{Mitchell1988-zr, George1993-au}. Thus, the spike-and-slab prior on the $j$th coefficient $\beta_j$ can be written as
\begin{align}
    p(\beta_j\vert \gamma_j, \nu^2) = \gamma_j N(0, \nu^2 \sigma_j^2) + (1-\gamma_j)\delta_0(\beta_j)
\end{align}
where $\delta_0(\cdot)$ is the Dirac measure (point mass at 0). The prior on $\gamma_j$ is given by
\begin{align}
    p(\gamma_j \vert \pi) = \pi^{\gamma_j}(1-\pi)^{1-\gamma_j}.
\end{align}
Letting $\odot$ denote elementwise multiplication, i.e., where $(a_1,  \ldots, a_n) \odot (b_1,  \ldots, b_n) = (a_1 b_1, \ldots, a_n b_n)$, the full pseudo-posterior when a spike-and-slab prior distribution is specified for $\betaVector$ can be written as
\begin{align}
    p(\betaVector, \lambdaVector, \ssParamVector \vert \mathbf{y}, \fullCovariateMatrix, \pi, \nu) 
    =& \prod_{i = 1}^n p(y_i \vert \betaVector, \ssParamVector, \lambdaVector) \prod_{j=1}^p \left[ p(\beta_j\vert \gamma_j, \nu^2) p(\gamma_j \vert \pi) \right] \nonumber \\
    =& \prod_{i=1}^n \frac{1}{\sqrt{2 \pi \lambda_i}} \exp\left\{\frac{1}{2} \frac{(1 + \lambda_i - y_i x_i^\top (\ssParamVector \odot \betaVector))^2}{\lambda_i}\right\} \nonumber  \\
    & \times \prod_{j=1}^p  \left[  (\gamma_j N(0, \nu^2 \sigma_j^2) + (1-\gamma_j)\delta_0(\beta_j)) \pi^{\gamma_j}(1-\pi)^{1-\gamma_j}\right].
\end{align}

\subsection{Estimation}\label{sec:estimation}
To draw from the pseudo-posterior distribution, \cite{Polson2011-rw} employed two algorithms, an expectation-minimization (EM) approach and a Gibbs sampling approach. The approach we take is the latter. Although sampling the pseudo-posterior is likely to be more time intensive than estimation via the EM algorithm, the rationale for a fully Bayesian approach is to enable uncertainty quantification (\cref{sec:uncertQuant}).

\subsubsection{Normal prior distribution for \texorpdfstring{$\betaVector$}{betaVector}}
The full pseudo-posterior distribution under normal priors for $\betaVector$ \cref{eq:full_pseudoposterior_normalprior} has two unknown parameters, $\betaVector$ and $\lambdaVector$. To sample these parameters, we derive a Gibbs sampling algorithm, which entails sequentially sampling each parameter conditionally on the most up-to-date values of the other parameters. We give a high level summary of the derivation in this section and full details in \cref{appendix2}. The conditional distribution of $\lambda_i \vert \betaVector,  \fullCovariateVectorObserved_i, \actionVectorObserved_i, \rewardVectorObserved_i$ (up to a normalizing constant) can be written as:
\begin{align}
    p(\lambda_i &\vert \betaVector, \fullCovariateVectorObserved_i, \actionObserved_i, \rewardObserved_i) 
    \times \mathbbm{1}(\actionObserved = -1) \lambda_i^{-1/2} \cdot  \exp\left\{-\frac{1}{2}  \left(\lambda_i + \left(\frac{\rewardObserved_i}{1-\rho} \right)^2 (1 - \actionObserved_i \fullCovariateVectorObserved_i^\top \betaVector) \lambda_i^{-1} \right)  \right\}.  \nonumber 
\end{align}
From \cite{Devroye1986-aj}, page 479, a random variable has the generalized inverse Gaussian distribution $\mathcal{GIG}(\gamma, \psi, \chi)$ if its density function is
$p(x \vert \gamma, \psi, \chi) = C(\gamma, \psi, \chi) x^{\gamma - 1} \exp \left\{ -\frac{1}{2} \left(\frac{\chi}{x} + \psi x \right) \right\}$, where $C(\gamma, \psi, \chi)$ is a normalization constant. Thus 
\begin{align}\label{eq:conditionalForLambda}
    p(\lambda_i &\vert \betaVector, \fullCovariateVectorObserved_i, \actionObserved_i, \rewardObserved_i) 
    \sim \mathbbm{1}(\actionObserved_i = 1) \mathcal{GIG}\left(\frac{1}{2}, 1, \left(\frac{\rewardObserved_i}{\rho}\right)^2(1 - \actionObserved_i \fullCovariateVectorObserved_i^\top \betaVector)^2\right) \nonumber \\
    &+ \mathbbm{1}(\actionObserved_i = -1) \mathcal{GIG}\left(\frac{1}{2}, 1, \left(\frac{\rewardObserved_i}{1 - \rho}\right)^2(1 - \actionObserved_i \fullCovariateVectorObserved_i^\top \betaVector)^2\right).
\end{align}

The conditional distribution of $\betaVector \vert \lambdaVector, \omegaVector, \rewardVectorObserved, \actionVectorObserved, \fullCovariateVectorObserved$ follows from standard arguments for Bayesian linear models. The notable difference from such a standard model is that $\betaVector \vert \lambdaVector, \omegaVector, \rewardVectorObserved, \actionVectorObserved, \fullCovariateVectorObserved$ is a mixture over two distributions, one for when the observed treatment  in the data under analysis is $1$ and one for when the observed treatment is $-1$. 
The conditional distribution of $\betaVector \vert \lambdaVector, \rewardVectorObserved, \actionVectorObserved, \fullCovariateVectorObserved$ has the form
\begin{align}
    p(\betaVector &\vert \lambdaVector, \rewardVectorObserved, \actionVectorObserved, \fullCovariateVectorObserved) \nonumber \\
    \propto & \exp \left\{ -\frac{1}{2} \sum_{\{i: \actionObserved_i = 1\}} \left(-2 \frac{\rewardObserved_i}{\propensityScore}\actionObserved_i \fullCovariateVectorObserved_i^\top \betaVector \left(1 + \frac{\rewardObserved_i}{\propensityScore \lambda_i} \right) +  \left(\frac{\rewardObserved_i}{\propensityScore} \right)^2 \frac{1}{\lambda_i} (\actionObserved_i \fullCovariateVectorObserved_i^\top \betaVector)^2  \right) \right\} \nonumber \\
    & \cdot \exp \left\{ -\frac{1}{2} \sum_{\{i: \actionObserved_i = -1\}} \left(  -2 \frac{\rewardObserved_i}{1 - \propensityScore} \actionObserved_i \fullCovariateVectorObserved_i^\top \betaVector \left(1 + \frac{1}{(1 - \propensityScore) \lambda_i} \right) + \left(\frac{\rewardObserved_i}{1 - \propensityScore} \right)^2 \frac{1}{\lambda_i}  (\actionObserved_i \fullCovariateVectorObserved_i^\top \betaVector)^2   \right) \right\} \exp\left\{-\frac{1}{2} \sum_{j = 1}^p \frac{(\beta_j - \mu_{0,1})^2}{\sigma_0^2 } \right\} \nonumber
\end{align}
Let $n_{1} = \sum_{i = 1}^n \mathbbm{1}(\actionObserved_i = 1)$ and $n_{-1} = \sum_{i = 1}^n \mathbbm{1}(\actionObserved_i = -1)$. Define $\matrixSymbol{X}_1$, $\matrixSymbol{W}_1$, $\matrixSymbol{R}_1$, and $\matrixSymbol{\Lambda}_1$ as
\begin{align}
    \matrixSymbol{X}_1 &\equiv 
    \begin{pmatrix}
          \actionObserved_1 x_{1,1}    & \cdots    & \actionObserved_1 x_{1,p} \\
         \vdots     &           & \vdots \\
          \actionObserved_{n_1} x_{n_1, 1} & \cdots    & \actionObserved_{n_1} x_{n_1, p}
    \end{pmatrix}_{(n_1 \times p)}, \qquad 
    &\matrixSymbol{W}_1 \equiv
    \begin{pmatrix}
        1 + \frac{\rewardObserved_1}{\lambda_1} \\
        \vdots \\
        1 + \frac{\rewardObserved_{n_1}}{\lambda_{n_1}}
    \end{pmatrix}_{(n_1 \times 1)},  \nonumber \\ 
    \matrixSymbol{R}_1 &\equiv diag(r_1/\propensityScore, \ldots, r_{n_1/\propensityScore})_{(n_1 \times n_1)}, \text{ and }
    &\matrixSymbol{\Lambda}_1 = diag(\lambda_1, \ldots, \lambda_{n_1}). 
\end{align}
Define $\matrixSymbol{X}_{-1}$, $\matrixSymbol{W}_{-1}$, $\matrixSymbol{R}_{-1}$, and $\boldsymbol{\Lambda}_{-1}$ analogously. Additionally define and $\matrixSymbol{\Sigma}$ as $diag(\sigma_1, \ldots, \sigma_p)$.

Then, we have that
\begin{align}
   p(\betaVector &\vert \lambdaVector, \omegaVector, \rewardVectorObserved, \actionVectorObserved, \fullCovariateVectorObserved) \nonumber \\
    \propto& \exp \bigg\{-\frac{1}{2} \Bigg[\betaVector^\top \underbrace{\left(\matrixSymbol{X}_1^\top \matrixSymbol{R}_1^\top \matrixSymbol{\Lambda}_1^{-1} \matrixSymbol{R}_1 \matrixSymbol{X}_1 + \matrixSymbol{X}_{-1}^\top \matrixSymbol{R}_{-1}^\top \matrixSymbol{\Lambda}_{-1}^{-1} \matrixSymbol{R}_{-1} \matrixSymbol{X}_{-1} +  \matrixSymbol{\Sigma}^{-1}\right)}_{\equiv B_1^{-1}}  \betaVector \nonumber \\
    & -2 (\underbrace{\matrixSymbol{W}_1^\top \matrixSymbol{R}_1\matrixSymbol{X}_1 + \matrixSymbol{W}_{-1}^\top \matrixSymbol{R}_{-1} \matrixSymbol{X}_{-1} + \betaPriorMean^\top \matrixSymbol{\Sigma}^{-1} }_{\equiv b_1})\betaVector \Bigg] \bigg\} \nonumber \\
    \propto& \exp \left\{-\frac{1}{2} (\betaVector - B_1 b_1)^\top B_1^{-1} (\betaVector - B_1 b_1) \right\}. \nonumber
\end{align}
In other words, the conditional distribution of $\betaVector$ given $\lambdaVector$, $\omegaVector$, and is multivariate normal with mean $B_1 b_1$ and variance-covariance matrix $B_1$. With the necessarily conditional distributions derived, the Gibbs sampling algorithm for sampling from the posterior distribution is given in Box 1. 

\begin{mybox}
    \begin{center}
        \textbf{Box 1.} Gibbs sampling algorithm for normal distribution priors on $\betaVector$
    \end{center}
    Initialize $\lambdaVector$ and $\betaVector$; set the hyperparameters $\betaPriorMean$ and $\betaPriorVar$.
    
    \textbf{Step 1}: Draw $\betaVector^{(g+1)} \vert \lambdaVector^{(g)}, \rewardVectorObserved, \actionVectorObserved, \fullCovariateVectorObserved \sim \mathcal{N}(B_1^{(g)}b_1^{(g)}, B_1^{(g)})$.

    \textbf{Step 2}: Draw $\lambdaVector^{-1(g+1)} \vert \betaVector^{(g)}, \rewardVectorObserved, \actionVectorObserved, \fullCovariateVectorObserved$ where
    \begin{align*}
        \lambda_i \sim  \mathbbm{1}(\actionObserved_i = 1) \mathcal{GIG}\left(\frac{1}{2}, 1, \left(\frac{\rewardObserved_i}{\rho}\right)^2(1 - \actionObserved_i \fullCovariateVectorObserved_i^\top \betaVector)^2\right) 
    + \mathbbm{1}(\actionObserved_i = -1) \mathcal{GIG}\left(\frac{1}{2}, 1, \left(\frac{\rewardObserved_i}{1 - \rho}\right)^2(1 - \actionObserved_i \fullCovariateVectorObserved_i^\top \betaVector)^2\right).
    \end{align*}
    Repeat Steps 1 and 2 until the chains converge.
\end{mybox}

\subsubsection{Exponential power prior distribution for \texorpdfstring{$\betaVector$}{betaVector}}

The full pseudo-posterior distribution when an exponential power prior distribution is specified for $\betaVector$ (\cref{eq:full_pseudoposterior_exponentialPowerPrior}) has three unknown parameters, $\betaVector$, $\lambdaVector$, and $\omegaVector$. To sample these parameters, we derive a Gibbs sampling algorithm, which entails sequentially sampling each parameter conditionally on the most up-to-date values of the other parameters. We give a high level summary of the derivation in this section and full details in \cref{appendix2}. The conditional distribution of $\lambda_i \vert \betaVector,  \fullCovariateVectorObserved_i, \actionVectorObserved_i, \rewardVectorObserved_i$ is the same as in the case with normal prior distributions for $\betaVector$ given in \cref{eq:conditionalForLambda}.

The conditional distribution of $\betaVector \vert \lambdaVector, \omegaVector, \rewardVectorObserved, \actionVectorObserved, \fullCovariateVectorObserved$ follows from standard arguments for Bayesian linear models. The notable difference from such a standard model is that $\betaVector \vert \lambdaVector, \omegaVector, \rewardVectorObserved, \actionVectorObserved, \fullCovariateVectorObserved$ is a mixture over two distributions, one for when the observed treatment  in the data under analysis is $1$ and one for when the observed treatment is $-1$. The third term of the conditional distribution is the penalty. The conditional distribution of $\betaVector \vert \lambdaVector, \omegaVector, \rewardVectorObserved, \actionVectorObserved, \fullCovariateVectorObserved$ has the form
\begin{align}
    p(\betaVector &\vert \lambdaVector, \omegaVector, \rewardVectorObserved, \actionVectorObserved, \fullCovariateVectorObserved) \nonumber \\
    \propto & \exp \left\{ -\frac{1}{2} \sum_{\{i: \actionObserved_i = 1\}} \left(-2 \frac{\rewardObserved_i}{\propensityScore}\actionObserved_i \fullCovariateVectorObserved_i^\top \betaVector \left(1 + \frac{\rewardObserved_i}{\propensityScore \lambda_i} \right) +  \left(\frac{\rewardObserved_i}{\propensityScore} \right)^2 \frac{1}{\lambda_i} (\actionObserved_i \fullCovariateVectorObserved_i^\top \betaVector)^2  \right) \right\} \nonumber \\
    & \cdot \exp \left\{ -\frac{1}{2} \sum_{\{i: \actionObserved_i = -1\}} \left(  -2 \frac{\rewardObserved_i}{1 - \propensityScore} \actionObserved_i \fullCovariateVectorObserved_i^\top \betaVector \left(1 + \frac{1}{(1 - \propensityScore) \lambda_i} \right) + \left(\frac{\rewardObserved_i}{1 - \propensityScore} \right)^2 \frac{1}{\lambda_i}  (\actionObserved_i \fullCovariateVectorObserved_i^\top \betaVector)^2   \right) \right\} \exp\left\{-\frac{1}{2 \nu^2} \sum_{j = 1}^p \frac{\beta_j^2}{\sigma_j^2 \omega_j} \right\}. \nonumber
\end{align}
Letting $ \matrixSymbol{\Omega} \equiv diag(\omega_1, \ldots, \omega_p)_{(p \times p)}$, we have that
\begin{align}
   p(\betaVector &\vert \lambdaVector, \omegaVector, \rewardVectorObserved, \actionVectorObserved, \fullCovariateVectorObserved) \nonumber \\
    \propto& \exp \bigg\{-\frac{1}{2} \Bigg[\betaVector^\top \underbrace{\left(\matrixSymbol{X}_1^\top \matrixSymbol{R}_1^\top \matrixSymbol{\Lambda}_1^{-1} \matrixSymbol{R}_1 \matrixSymbol{X}_1 + \matrixSymbol{X}_{-1}^\top \matrixSymbol{R}_{-1}^\top \matrixSymbol{\Lambda}_{-1}^{-1} \matrixSymbol{R}_{-1} \matrixSymbol{X}_{-1} + \nu^{-2} \matrixSymbol{\Omega}^{-1} \matrixSymbol{\Sigma}^{-1}\right)}_{\equiv B_2^{-1}}  \betaVector 
     -2 (\underbrace{\matrixSymbol{W}_1^\top \matrixSymbol{R}_1\matrixSymbol{X}_1 + \matrixSymbol{W}_{-1}^\top \matrixSymbol{R}_{-1} \matrixSymbol{X}_{-1} }_{\equiv b_2})\betaVector \Bigg] \bigg\} \nonumber \\
    \propto& \exp \left\{-\frac{1}{2} (\betaVector - B_2 b_2)^\top B_2^{-1} (\betaVector - B_2 b_2) \right\}. \nonumber
\end{align}
Thus, the conditional distribution of $\betaVector$ given $\lambdaVector$, $\omegaVector$, and is multivariate normal with mean $B_2 b_2$ and variance-covariance matrix $B_2$. Full details of this derivation are given in \cref{appendix2}.

Finally, the conditional distribution of $\omegaVector \vert \betaVector, \nu$ is the same as that given in Corollary 3 of \Citet{Polson2011-rw}. For $\alpha = 1$, the full conditional distribution of $\omega$ is $\omega_j^{-1} \vert \beta_j, \nu \sim \mathcal{IG}(\nu \sigma_j/\vert \beta_j \vert, 1)$. Together, with these three conditional distributions, we can summarize the Gibbs sampling algorithm as in Box 2. 

\begin{mybox}
    \begin{center}
        \textbf{Box 2.} Gibbs sampling algorithm for exponential power distribution prior on $\betaVector$
    \end{center}
    Initialize $\lambdaVector$, $\betaVector$ and $\omegaVector$.
    
    \textbf{Step 1}: Draw $\betaVector^{(g+1)} \vert \nu, \matrixSymbol{\Lambda}^{(g)}, \matrixSymbol{\Omega}^{(g)}, \rewardVectorObserved, \actionVectorObserved, \fullCovariateVectorObserved \sim \mathcal{N}(B_2^{(g)}b_2^{(g)}, B_2^{(g)})$.

    \textbf{Step 2}: Draw $\lambdaVector^{-1(g+1)} \vert \betaVector^{(g+1)}, \rewardVectorObserved, \actionVectorObserved, \fullCovariateVectorObserved$ where
    \begin{align*}
        \lambda_i^{(g+1)} \vert \betaVector^{(g+1)}, \nu, r_i, x_i \sim&  \mathbbm{1}(\actionObserved_i = 1) \mathcal{GIG}\left(\frac{1}{2}, 1, \left(\frac{\rewardObserved_i}{\rho}\right)^2(1 - \actionObserved_i \fullCovariateVectorObserved_i^\top \betaVector^{(g+1)})^2\right) \\
    &+ \mathbbm{1}(\actionObserved_i = -1) \mathcal{GIG}\left(\frac{1}{2}, 1, \left(\frac{\rewardObserved_i}{1 - \rho}\right)^2(1 - \actionObserved_i \fullCovariateVectorObserved_i^\top \betaVector^{(g+1)})^2\right).
    \end{align*}
    
    \textbf{Step 3}: Draw $\omega_j^{-1(g+1)} \vert \beta_j^{(g+1)}, \nu \sim \mathcal{IG}(\nu \sigma_j \vert \beta_j \vert^{-1}, 1)$
    
    Repeat Steps 1, 2, and 3 until the chains converge.
\end{mybox}

\subsubsection{Spike-and-slab prior distribution for \texorpdfstring{$\betaVector$}{betaVector}}

The full pseudo-posterior distribution when a spike-and-slab prior is specified for $\betaVector$ has three unknown parameters, $\betaVector$, $\lambdaVector$, and $\gammaVector$. Similarly to the procedure outlined earlier, we employ a Gibbs sampling algorithm to obtain these parameters. The conditional distribution of $\lambda_i \vert \betaVector,  \fullCovariateVectorObserved_i, \actionVectorObserved_i, \rewardVectorObserved_i$ is the same as in the case with normal prior distributions for $\betaVector$ given in \cref{eq:conditionalForLambda}. 

The conditional distribution of $\betaVector$ given $\lambdaVector$, $\rewardVectorObserved$, $\actionVectorObserved$, and $\fullCovariateVectorObserved$ mirrors the previous exponential power prior distribution, but without the third term of the penalty term. The adjustments are slight, with only the following modifications:  

\begin{align*}
    B^{-1}_{\gamma} &=  X^T_{1} R_{1}^T \Lambda^{-1}_{1} R_1 X_1 + X^T_{-1} R_{-1}^T \Lambda^{-1}_{-1} R_{-1} X_{-1} \\
    b_{\gamma} &= B_{\gamma} (W^T_{1}R_{1}X_{1} + W^T_{-1}R_{-1}X_{-1})
\end{align*}

The spike-and-slab prior induces sparsity in coefficients through the parameters $\mathbf{\gamma}$. The conditional distribution of $\gammaVector$ given $\lambdaVector$, $\rewardVectorObserved$, $\actionVectorObserved$, and $\fullCovariateVectorObserved$ may be written as in \Citet{Polson2011-rw} with $b_{\gamma}$ and $B_{\gamma}$ previously introduced :

\begin{align*}
    p(\gamma | \gammaVector, \rewardVectorObserved, \actionVectorObserved, \fullCovariateVectorObserved,  \nu) \propto p(\gamma) \frac{| \Sigma^{-1}_{\gamma} / \nu^2|^{1/2}} {|B_{\gamma}^{-1}|^{1/2}} \exp{(-\frac{1}{2} \sum_{i=1}^n \frac{(1+\lambda_i - a_i x_{i,\gamma}^T b_{\gamma})^2}{\lambda_i} - \frac{1}{2\nu^2} b_{\gamma}^T \Sigma^{-1}_{\gamma} b_{\gamma} )}
\end{align*}

By exploiting the quadratic term, the conditional distribution can be rewritten as in step 2 of Box 3. This form includes static terms that do not need to be recomputed in every iteration. To sample $\gammaVector$, a second Gibbs sampler nested within the first must be implemented. 

\begin{mybox}
    \begin{center}
        \textbf{Box 3.} Gibbs sampling algorithm for spike-and-slab prior distribution on $\betaVector$
    \end{center}
    Initialize $\betaVector$ and $\gammaVector$. 
    
    \textbf{Step 1}: Draw $\lambdaVector^{-1(g+1)} \vert \betaVector^{(g+1)}, \rewardVectorObserved, \actionVectorObserved, \fullCovariateVectorObserved$ where
    \begin{align*}
        \lambda_i^{(g+1)} \vert \betaVector^{(g+1)}, \nu, r_i, x_i \sim&  \mathbbm{1}(\actionObserved_i = 1) \mathcal{GIG}\left(\frac{1}{2}, 1, \left(\frac{\rewardObserved_i}{\rho}\right)^2(1 - \actionObserved_i \fullCovariateVectorObserved_i^\top \betaVector^{(g+1)})^2\right) \\
    &+ \mathbbm{1}(\actionObserved_i = -1) \mathcal{GIG}\left(\frac{1}{2}, 1, \left(\frac{\rewardObserved_i}{1 - \rho}\right)^2(1 - \actionObserved_i \fullCovariateVectorObserved_i^\top \betaVector^{(g+1)})^2\right).
    \end{align*}

    \textbf{Step 2}: For $i=1,\dots,k$ draw $\gamma_i$ from $p(\gamma_i | \gamma_{-i})$ which is proportional to 
    \begin{align*}
        p(\gamma | \gammaVector, \rewardVectorObserved, \actionVectorObserved, \fullCovariateVectorObserved,  \nu) \propto \\
        p(\gamma) \frac{| \Sigma^{-1}_{\gamma} / \nu^2|^{1/2}} {|B_{\gamma}^{-1}|^{1/2}} \exp{(-\frac{1}{2} [c(\lambda) + b_{\gamma}^T (X^T \Lambda^{-1} X)_{\gamma} b_{\gamma} - 2b_{\gamma}^T [X^T(1+\lambda^{-1})]_{\gamma}] - \frac{1}{2\nu^2} b_{\gamma}^T \Sigma^{-1}_{\gamma} b_{\gamma} )}
    \end{align*}
    
    \textbf{Step 3}: When $\gamma_i = 1$, draw $\betaVector^{(g+1)}_{\gamma} \vert \nu, \matrixSymbol{\Lambda}^{(g)}, \rewardVectorObserved, \actionVectorObserved, \fullCovariateVectorObserved \sim \mathcal{N}(b_{\gamma}^{(g)}, B_{\gamma}^{(g)})$.
    
    Repeat Steps 1, 2, and 3 until the chains converge.
\end{mybox}

\subsection{Prediction and uncertainty quantification}\label{sec:uncertQuant}
Using the posterior predictive distribution, we can make treatment recommendations for a new patient and quantify our uncertainty in our recommendation. Let $\Theta = \{\betaVector, \lambdaVector\}$ and $\newAction$ denote the recommended treatment for a new patient with features $\newCovariate$. Then
\begin{align}
    p(\newAction = 1 \vert \newCovariate, \fullCovariateMatrix, \rewardVectorObserved, \actionVectorObserved) = \int_{\Theta} p(\newAction = 1 \vert \newCovariate, \fullCovariateMatrix, \rewardVectorObserved, \actionVectorObserved, \betaVector, \lambdaVector) p(\betaVector, \lambdaVector\vert \fullCovariateVectorObserved, \rewardVectorObserved, \actionVectorObserved) d\theta. \nonumber
\end{align}
For the class predictions, we can use the probit model which has the form
\begin{align}
    p(a = 1 \vert x) = \Phi(x^\top\beta) \nonumber
\end{align}
where $\Phi$ is the cumulative distribution function of the standard normal distribution. Thus we can write the posterior predictive distribution as 
\begin{align}
    p(\newAction = 1 \vert \newCovariate, \fullCovariateMatrix, \rewardVectorObserved, \actionVectorObserved) = \int_{\Theta} \Phi(\newCovariate^\top \betaVector) p(\betaVector, \lambdaVector\vert \fullCovariateVectorObserved, \rewardVectorObserved, \actionVectorObserved) d\theta.
\end{align}

\section{Simulation study}\label{sec:simulation}

\subsection{Classification performance}\label{sec:classification}

We conducted simulation studies to assess the classification performance of the proposed method, following \cite{Zhao2012-vc} and \cite{Song2015-yk}. We compared the performance of OWL, Bayesian OWL with normal priors for $\betaVector$, Bayesian OWL with exponential power prior for $\betaVector$, and Bayesian OWL with spike-and-slab prior for $\betaVector$. For each simulated patient, we generated a 10-dimensional vector of patient features, $X_1, \ldots, X_{10}$, drawn independently and uniformly distributed on $[-1,1]$. Treatment $A$ was drawn from $\{-1,1\}$ independently of the prognostic variables with $\mathbb{P}(A=1) = 1/2$. The outcome variable $R$ was normally distributed with mean $Q_0 = 1 + 2X_1 + X_2 + 0.5X_3 + T_0(X,A)$ and standard deviation 1, where $T_0(X,A)$ was the interaction term between treatment and patient features. We examined two scenarios for the treatment-feature interaction term: 
\begin{itemize}
    \item Scenario 1 : $T_0(A,X) = (X_1 + X_2)A$
    \item Scenario 2 : $T_0(A,X) =0.442(1-X_1 -X_2)A$
\end{itemize}

Both scenarios 1 and 2 had linear decision boundaries determined by $X_1$ and $X_2$. For scenario 1, the true optimal rule was given by $\mathbb{1}(X_1 + X_2 > 0)$, while for scenario 2, it was $\mathbb{1}(1 - X_1 - X_2 > 0)$. OWL was implemented with a linear kernel. For Bayesian OWL, Gibbs sampling was used to draw from the posterior distributions of the parameters 500 times. The first 150 draws were discarded as "burn-in" and point estimates of $\betaVector$ were computed by taking the mean of the draws from the posterior distribution. Throughout, we set the hyperparameter $\nu = 0.8$.  

For each scenario, we varied the training dataset from 100 to 200, 400 and 800 and tested on 1000 patients. For each training set size, we conducted 200 simulation runs. We evaluated classification performance using the misclassification rate, the ratio of the number of patients recommended a treatment counter to the true optimal rule divided by the total number of patients in the simulation run ($\frac{\text{Number of patients misclassified}}{\text{Total number of patients}}$). The simulation results are presented in Table 1 and 2. 

\begin{table}[h]
\centering
\begin{tabular}{cccccc}
\toprule
& & Bayesian OWL & Bayesian OWL & Bayesian OWL\\
$n$ & OWL & Normal Prior &  Exponential Power Prior & Spike and Slab \\
\midrule
100 & 0.24 & 0.38 & 0.38 & 0.39 \\
200 & 0.18 & 0.34 & 0.34 & 0.34 \\
400 & 0.13 & 0.29 & 0.29 & 0.30 \\
800 & 0.10 & 0.24 & 0.24 & 0.26 \\
\bottomrule
\end{tabular}
\caption{Misclassification rates for different methods and sample sizes for scenario 1.}
\label{tab:results1}
\end{table}

\cellcolor{lightblue} 

\begin{table}[h]
\centering
\begin{tabular}{cccccc}
\toprule
& & Bayesian OWL & Bayesian OWL & Bayesian OWL \\
$n$ & OWL & Normal Prior &   Exponential Power Prior & Spike and Slab \\
\midrule
100 & 0.22 & 0.38 & 0.38 & 0.39 \\
200 & 0.15 & 0.34 & 0.34 & 0.34 \\
400 & 0.13 & 0.31 & 0.31 & 0.30 \\
800 & 0.10 & 0.25 & 0.25 & 0.22 \\
\bottomrule
\end{tabular}
\caption{Misclassification rates for different methods and sample sizes for scenario 2.}
\label{tab:results2}
\end{table}

As expected, the classification performance improved among all the ITR learning methods evaluated as the sample size increased. However, OWL consistently outperformed Bayesian OWL in all sample sizes and in both scenarios. We hypothesize that, with additional hyperparameter tuning, the performance of Bayesian OWL can be improved. Ordinarily, one would be hesitant to propose a method that is dominated by an existing method. However, the dominance of OWL is with respect to the misclassification rate. OWL, even with 800 samples in our simulation, has a 10\% misclassification rate, and there is no way to determine which of the 10\% of the simulated patients are likely misclassified (given a non-optimal treatment recommendation). In contrast, Bayesian OWL yields the entire posterior distribution of the estimated optimal ITR and thereby allows for immediate uncertainty quantification of individual-level treatment recommendations. In essence, Bayesian OWL can inform us of which treatment recommendations it is less certain about whereas OWL cannot. We demonstrate this in \cref{sec:simulation_uncertaintyQuantification}.   

\subsection{Treatment recommendation uncertainty quantification}\label{sec:simulation_uncertaintyQuantification}
To highlight the utility of quantifying the uncertainty of individual-level treatment recommendations, we generated a data set of 1000 patients under Scenario 1. We used these simulated data to train a Bayesian OWL model using the exponential power power prior. Next, we used the same generative approach to simulate another 1000 patients. Specifically, we used a fine grid to generate $X_1$ and $X_2$, the key variables in the true optimal rule (i.e., tailoring variables). The rationale for this approach was to generate a simulated set of patients whose characteristics covered the domain of the true optimal ITR so that we could estimate the uncertainty for combinations of $X_1$ and $X_2$ throughout the domain, $\mathcal{X}_1 \times \mathcal{X}_2 \in [-1,1]^2$. 

\begin{figure}[ht]
  \centering
  \includegraphics[width=0.9\textwidth]{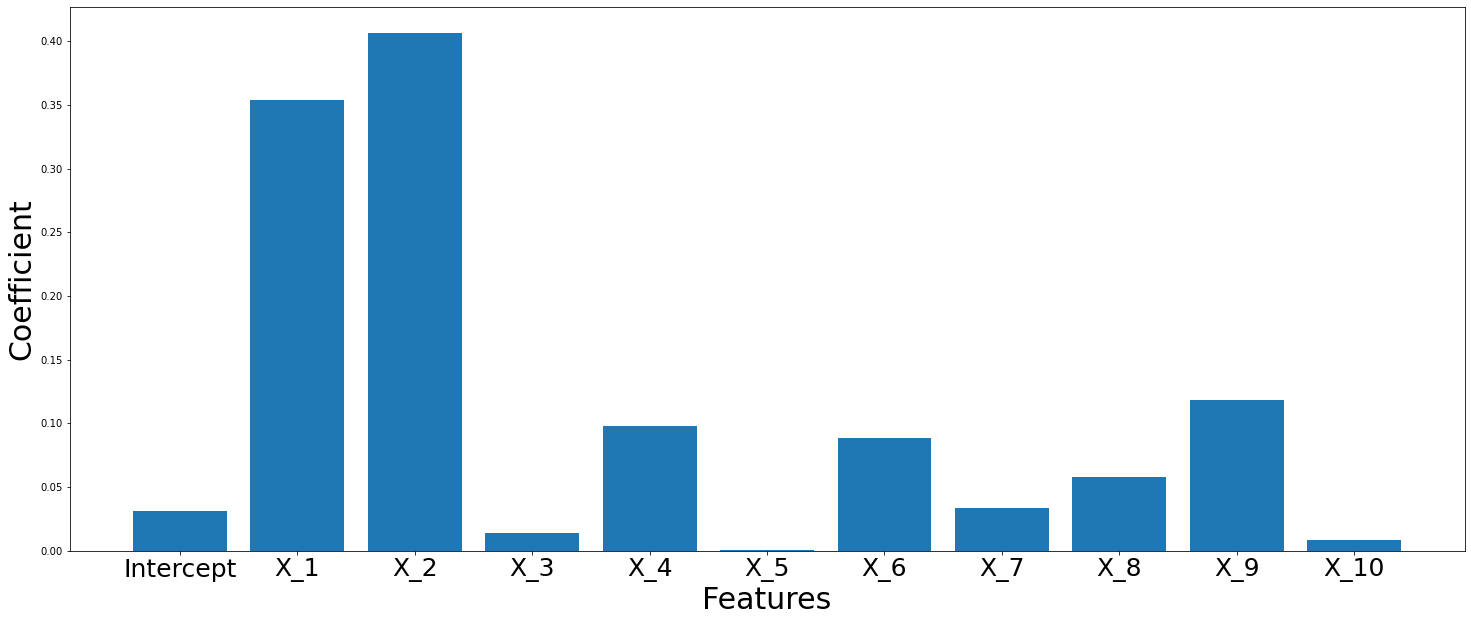}
  \caption{Feature importance}
  \label{fig:features}
\end{figure}

We evaluated the coefficients for the trained Bayesian OWL model. With an exponential power prior on $\betaVector$, which is analogous to $L1$ regularization, we would expect the coefficients of the features in the true optimal rule (tailoring variables) to be large and the coefficients of the features not in the rule to be driven close to 0. The magnitudes of the coefficients are displayed in \cref{fig:features}. As we would hope based on our knowledge of the true optimal rule, the magnitudes of the coefficients for $X_1$ and $X_2$ were larger than those for the other features, indicating that the estimated ITR using Bayesian OWL made decisions based on the correct patient features. 
 
\begin{figure}[ht]
  \centering  \includegraphics[width=0.8\textwidth]{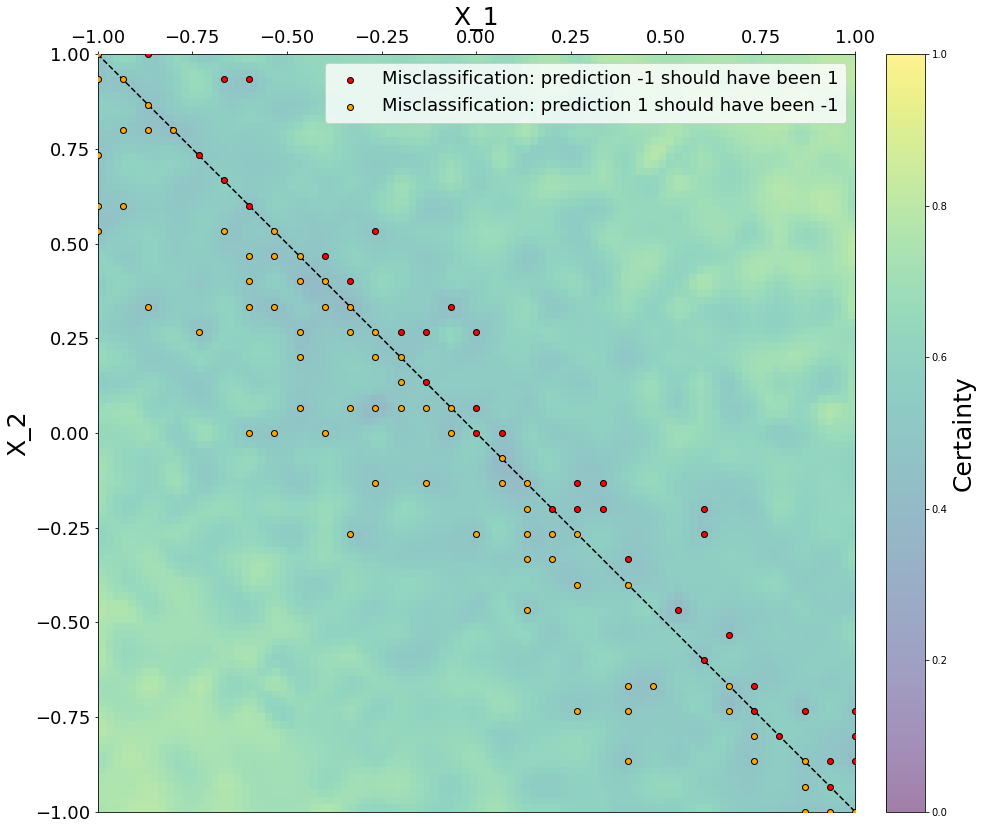}
  \caption{Heatmap of uncertainty quantification}
  \label{fig:heatmap}
\end{figure}

\cref{fig:heatmap} demonstrates Bayesian OWL's ability to quantify uncertainty in its treatment recommedations. In Scenario 1, the true optimal ITR divides patients into two groups: patients in the upper-right half of the graph (where $X_1 + X_2 > 0$) should ideally get treatment $A = 1$, while those in the lower-left half should get treatment $A = -1$. Using the posterior predictive distribution as in \cref{sec:uncertQuant}, we computed the uncertainty associated with recommending Treatment 1 for patients whose features $X_1$ and $X_2$ lie in the upper-right half of the graph and the uncertainty associated with recommending Treatment -1 in the lower-left half. Notably, uncertainty was evaluated \textit{individually} for each of the 1000 patients in our test cohort. This means that for each individual, we estimated how certain or uncertain we were about their specific treatment recommendation given their features $X_1$ and $X_2$. Moreover, because our test set included simulated patients whose features spanned the domain of the ITR, we were able to compute uncertainty for every ``type" of patient who could be recommended a treatment using the estimated ITR. We visualize the uncertainty across the domain of the ITR using a heat map (\cref{fig:heatmap}). Certainties close to 1 (less uncertainty) are lighter in color and depicted with yellow and light green. In contrast, certainties close to 0 (more uncertain) are darker in color and depicted with purple and dark blue. As expected, Bayesian OWL is more certain about treatment recommendations for patients whose features are far from the decision boundary than for those that are close to the decision boundary. 

Furthermore, in \cref{fig:heatmap}, we have included misclassified individuals in our simulation. Those who were recommended treatment $-1$ but should have been recommended treatment $1$ are indicated by red points, and those who were recommended treatment $1$ but should have been recommended treatment $-1$ are indicated by orange points. We observe that the misclassified patients are located near the boundary, as we expect, and most noteworthy that they are located in regains where the model exhibits the greatest uncertainty (regions in which the background is shaded purple). 

\section{Discussion}\label{sec:discussion}
In this paper, we introduced a Bayesian formulation of OWL. To our knowledge, this is the first Bayesian strategy for \textit{directly} learning an ITR. Moreover, we demonstrate how the Bayesian approach can enable us to quantify the uncertainty in our treatment recommendations. Both tasks, learning the optimal ITR and uncertainty quantification, can be implemented through a simple Gibbs sampling strategy for sampling the posterior distribution. 

One may wonder why a Bayesian approach to OWL is needed since we already have OWL which can rely on methods from convex optimization. We believe that the answer to this question is two-fold. First, OWL is an appealing strategy for learning optimal ITRs because it models the decision rule directly rather than modeling conditional mean models and backing out the optimal ITR. Second, our Bayesian formulation of OWL yields a full pseudo-posterior distribution which means that we can quantify uncertainty in ITR's treatment recommendations \textit{at the individual level}. This is in contrast to the more common approach in the ITR literature which involves estimating the uncertainty in the value of the proposed ITR, i.e., the uncertainty in the expected value we would observe if everyone in the population were treated according to the rule. This may prove useful in the design and implementation of clinical studies by providing a strategy for identifying the types of patients for whom we feel confident in our ability to make treatment recommendations and the patient types that may require additional sampling and information to improve the recommendations. By casting the direct ITR learning approach into a probabilistic framework, we have widened the inferential possibilities for directly learned ITRs. 

Our work has limitations. For example, we only examine linear rules. Although linear rules are simple to understand, there may be times when a nonlinear rule is desired or a nonlinear rule significantly outperforms a linear rule, i.e., clinically meaningful improvement from the nonlinear rule is worth the decrease in interpretability. \cite{Henao2014-mg} proposed a strategy for Bayesian SVM for nonlinear decision boundaries, which is likely a good blueprint for extending this work to the nonlinear rule setting. Moreover, our approach does not attempt to do variable selection. This limits its applicability in high dimensional settings in which there is no information or weak information as to which tailoring covariates should be included in the treatment rule. \cite{Polson2011-rw} considers L1-regularization on the decision boundary coefficients as well as a spike-and-slab prior to induce sparsity. These may be reasonable strategies to approximate the penalized version of OWL introduced by \cite{Song2015-yk}. Finally, in simulation studies, OWL outperforms Bayesian OWL with respect to the misclassification rate, which is only ameliorated by the fact that Bayesian OWL can tell us when it is uncertain about its recommendations whereas OWL cannot.

While the direct learning, or classification, approach to learning optimal ITRs is incredibly powerful, leveraging tools from machine learning, inference and uncertainty quantification at the individual treatment recommendation level continues to be a challenge. Bayesian OWL overcomes this limitation by fully leveraging the benefits of direct-learning and the use of a probabilistic framework. Generating precision medicine evidence with wider inferential potential can improve our ability to build trust in these treatment algorithms and ultimately improve how we deploy precision medicine evidence in real-world health care decision making.

\bibliographystyle{unsrtnat}
\bibliography{references}  

\section{Appendix}
\subsection{Representation of the pseudolikelihood function}\label{appendix1}

Because we have assumed that the reward is is strictly positive, the weight $\frac{\rewardObserved_i}{\actionObserved_i \propensityScore + (1 - \actionObserved_i)/2}$ is also positive and can be brought inside of the maximization operator so that
\begin{align}
    L_i(\actionObserved_i \vert& \rewardObserved_i, \fullCovariateVectorObserved_i, \betaVector) \nonumber \\
    =& \exp\left\{ -2 \frac{\rewardObserved_i}{\actionObserved_i \propensityScore + (1 - \actionObserved_i)/2} \max(1 - \actionObserved_i \fullCovariateVectorObserved_i^\top \betaVector, 0)\right\} \nonumber \\
    =& \exp\left\{ -2 \max\left( \frac{\rewardObserved_i}{\actionObserved_i \propensityScore + (1 - \actionObserved_i)/2} (1 - \actionObserved_i \fullCovariateVectorObserved_i^\top \betaVector, 0)\right)\right\} \nonumber \\
    =& \mathbbm{1}(\actionObserved_i = 1)\exp\left\{ -2 \max\left( \frac{\rewardObserved_i}{\propensityScore} (1 - \actionObserved_i \fullCovariateVectorObserved_i^\top \betaVector, 0)\right)\right\} + 
    \mathbbm{1}(\actionObserved_i = -1)\exp\left\{ -2 \max\left( \frac{\rewardObserved_i}{1-\propensityScore} (1 - \actionObserved_i \fullCovariateVectorObserved_i^\top \betaVector, 0)\right)\right\}. \nonumber
\end{align}
The derivation of the pseudolikelihood representation follows \Citet{Polson2011-rw}: Andrews and Mallows (1974) showed that $\int_0^\infty \frac{a}{\sqrt{2 \pi \lambda}} e^{-\frac{1}{2} (a^2 \lambda + b^2 \lambda^{-1})} d\lambda = e^{-\vert ab \vert}$. Setting $a = 1$ and $b = u$, we have
\begin{align*}
    \int_0^\infty \frac{1}{\sqrt{2 \pi \lambda}} e^{-\frac{1}{2} (\lambda + u^2 \lambda^{-1})} d\lambda =& e^{-\vert u \vert}. \nonumber \\
\end{align*}
Multiplying through by $e^{-u}$ and recalling the identity $\max(u, 0) = \frac{1}{2}(\vert u \vert + u)$, we have
\begin{align*}
    e^{-u}\int_0^\infty \frac{1}{\sqrt{2 \pi \lambda}} e^{-\frac{1}{2} (\lambda + u^2 \lambda^{-1})} d\lambda =& e^{-\vert u \vert} e^{-u} \nonumber \\
    \implies \int_0^\infty \frac{1}{\sqrt{2 \pi \lambda}} e^{-\frac{1}{2} (\lambda + u^2 \lambda^{-1}) - u} d\lambda =& e^{-\vert u \vert -u} \nonumber \\
    \implies \int_0^\infty \frac{1}{\sqrt{2 \pi \lambda}} e^{-\frac{1}{2\lambda} (\lambda^2 + u^2 + 2u\lambda)} d\lambda =& e^{-\vert u \vert -u} \nonumber \\
    \implies \int_0^\infty \frac{1}{\sqrt{2 \pi \lambda}} e^{-\frac{1}{2\lambda} (u + \lambda)^2} d\lambda =& e^{-2 \max(u, 0)}. \nonumber \\
\end{align*}
Thus we can write the individual contribution of each observation to the marginal likelihood as
\begin{align*}
    L_i(\actionObserved_i \vert& \lambda_i, \rewardObserved_i, \fullCovariateVectorObserved_i, \betaVector) \nonumber \\
    =& \mathbbm{1}(\actionObserved_i = 1)\exp\left\{ -2 \max\left( \frac{\rewardObserved_i}{\propensityScore} (1 - \actionObserved_i \fullCovariateVectorObserved_i^\top \betaVector, 0)\right)\right\} + 
    \mathbbm{1}(\actionObserved_i = -1)\exp\left\{ -2 \max\left( \frac{\rewardObserved_i}{1-\propensityScore} (1 - \actionObserved_i \fullCovariateVectorObserved_i^\top \betaVector, 0)\right)\right\} \nonumber \\
    =& \mathbbm{1}(\actionObserved_i = 1) \int_{0}^{\infty} \frac{1}{\sqrt{2 \pi \lambda_i}} \exp \left\{-\frac{1}{2 \lambda_i} \left(\frac{\rewardObserved_i}{\propensityScore}(1 - \actionObserved_i \fullCovariateVectorObserved_i^\top \betaVector) +\lambda_i\right)^2 \right\} d\lambda_i  \nonumber\\
    &+ \mathbbm{1}(\actionObserved_i = -1) \int_{0}^{\infty} \frac{1}{\sqrt{2 \pi \lambda_i}} \exp \left\{-\frac{1}{2 \lambda_i} \left(\frac{\rewardObserved_i}{1-\propensityScore}(1 - \actionObserved_i \fullCovariateVectorObserved_i^\top \betaVector) +\lambda_i\right)^2 \right\} d\lambda_i \nonumber \\
    =& \mathbbm{1}(\actionObserved_i = 1) \int_{0}^{\infty} \frac{1}{ \sqrt{ 2 \pi \lambda_i}} \exp \left\{-\frac{1}{2 \lambda_i} \left(\frac{\rewardObserved_i}{\propensityScore} + \lambda_i - \frac{\rewardObserved_i}{\rho} \actionObserved_i \fullCovariateVectorObserved_i^\top \betaVector \right)^2 \right\} d\lambda_i  \nonumber\\
    &+ \mathbbm{1}(\actionObserved_i = -1) \int_{0}^{\infty} \frac{1}{\sqrt{2 \pi \lambda_i}} \exp \left\{-\frac{1}{2 \lambda_i} \left(\frac{\rewardObserved_i}{1-\propensityScore} +\lambda_i - \frac{\rewardObserved_i}{1-\propensityScore} \actionObserved_i \fullCovariateVectorObserved_i^\top \betaVector \right)^2 \right\} d\lambda_i, 
\end{align*}
and that $L_i(\actionObserved_i, \lambda_i \vert \rewardObserved_i, \fullCovariateVectorObserved_i, \betaVector)$ is a scale mixture of Gaussians.

\subsection{Derivation of the Gibbs sampling alogorithm}\label{appendix2}%

\subsubsection{Conditional distribution of \texorpdfstring{$\lambda_i \vert \betaVector,  \fullCovariateVectorObserved_i, \actionVectorObserved_i, \rewardVectorObserved_i$}{lambda given betaVector, fullCovariateVectorObserved, actionVectorObserved, rewardVectorObserved}}
\begin{align}
    p(\lambda_i &\vert \betaVector, \fullCovariateVectorObserved_i, \actionObserved_i, \rewardObserved_i) \nonumber \\
    &\propto  \mathbbm{1}(\actionObserved = 1) \lambda_i^{-1/2} \cdot  \exp\left\{-\frac{1}{2}  \frac{\left( \lambda_i - \frac{\rewardObserved_i}{\rho} (1 - \actionObserved_i \fullCovariateVectorObserved_i^\top \betaVector) \right)^2}{\lambda_i}  \right\}  \nonumber \\
    & \times \mathbbm{1}(\actionObserved = -1) \lambda_i^{-1/2} \cdot  \exp \left\{-\frac{1}{2} \frac{\left(\lambda_i - \frac{\rewardObserved_i}{1-\propensityScore} (1 - \actionObserved_i \fullCovariateVectorObserved_i^\top \betaVector) \right)^2}{\lambda_i}  \right\} \nonumber \\
    &= \mathbbm{1}(\actionObserved = 1) \lambda_i^{-1/2} \cdot  \exp\left\{-\frac{1}{2}  \frac{\left( \lambda_i^2 - 2 \lambda_i \left(\frac{r_i}{\rho}\right)  (1 - \actionObserved_i \fullCovariateVectorObserved_i^\top \betaVector) + \left(\frac{\rewardObserved_i}{\rho} \right)^2 (1 - \actionObserved_i \fullCovariateVectorObserved_i^\top \betaVector)^2\right)}{\lambda_i}  \right\}  \nonumber \\
    & \times \mathbbm{1}(\actionObserved = -1) \lambda_i^{-1/2} \cdot  \exp\left\{-\frac{1}{2}  \frac{\left( \lambda_i^2 - 2 \lambda_i \left(\frac{r_i}{1-\rho}\right)  (1 - \actionObserved_i \fullCovariateVectorObserved_i^\top \betaVector) + \left(\frac{\rewardObserved_i}{1-\rho} \right)^2 (1 - \actionObserved_i \fullCovariateVectorObserved_i^\top \betaVector)^2\right)}{\lambda_i}  \right\} \nonumber \\
    & \propto \mathbbm{1}(\actionObserved = 1) \lambda_i^{-1/2} \cdot  \exp\left\{-\frac{1}{2}  \left(\lambda_i + \left(\frac{\rewardObserved_i}{\rho} \right)^2 (1 - \actionObserved_i \fullCovariateVectorObserved_i^\top \betaVector) \lambda_i^{-1} \right)  \right\}  \nonumber \\
    & \times \mathbbm{1}(\actionObserved = -1) \lambda_i^{-1/2} \cdot  \exp\left\{-\frac{1}{2}  \left(\lambda_i + \left(\frac{\rewardObserved_i}{1-\rho} \right)^2 (1 - \actionObserved_i \fullCovariateVectorObserved_i^\top \betaVector) \lambda_i^{-1} \right)  \right\}  \nonumber 
\end{align}
From \cite{Devroye1986-aj}, page 479, a random variable has the generalized inverse Gaussian distribution $\mathcal{GIG}(\gamma, \psi, \chi)$ if its density function is
\begin{align*}
    p(x \vert \gamma, \psi, \chi) = C(\gamma, \psi, \chi) x^{\gamma - 1} \exp \left\{ -\frac{1}{2} \left(\frac{\chi}{x} + \psi x \right) \right\},
\end{align*}
where $C(\gamma, \psi, \chi)$ is a normalization constant. Thus 
\begin{align}
    p(\lambda_i &\vert \betaVector, \fullCovariateVectorObserved_i, \actionObserved_i, \rewardObserved_i) \nonumber \\
    &\sim \mathbbm{1}(\actionObserved_i = 1) \mathcal{GIG}\left(\frac{1}{2}, 1, \left(\frac{\rewardObserved_i}{\rho}\right)^2(1 - \actionObserved_i \fullCovariateVectorObserved_i^\top \betaVector)^2\right) + \mathbbm{1}(\actionObserved_i = -1) \mathcal{GIG}\left(\frac{1}{2}, 1, \left(\frac{\rewardObserved_i}{1 - \rho}\right)^2(1 - \actionObserved_i \fullCovariateVectorObserved_i^\top \betaVector)^2\right).
\end{align}
Recall that if $X \sim \mathcal{GIG}\left(\frac{1}{2}, \lambda, \chi \right)$, then $X^{-1} \sim \mathcal{IG}(\mu, \lambda)$ where $\chi = \lambda/\mu^2$ and $\mathcal{IG}$ denotes the inverse Gaussian distribution. Consequently, we can write the 
\begin{align}
     p(\lambda_i &\vert \betaVector, \fullCovariateVectorObserved_i, \actionObserved_i, \rewardObserved_i) \nonumber \\
    &\sim \mathbbm{1}(\actionObserved_i = 1) \mathcal{GIG}\left(\frac{1}{2}, 1, \left(\frac{\rewardObserved_i}{\rho}\right)^2(1 - \actionObserved_i \fullCovariateVectorObserved_i^\top \betaVector)^2\right) + \mathbbm{1}(\actionObserved_i = -1) \mathcal{GIG}\left(\frac{1}{2}, 1, \left(\frac{\rewardObserved_i}{1 - \rho}\right)^2(1 - \actionObserved_i \fullCovariateVectorObserved_i^\top \betaVector)^2\right).
\end{align}

\subsubsection{Conditional distribution of \texorpdfstring{$\betaVector \vert \lambdaVector,  \betaPriorMean, \betaPriorVar, \rewardVectorObserved, \actionVectorObserved, \fullCovariateVectorObserved$}{betaVector given lambdaVector betaPriorMean betaPriorVar rewardVectorObserved actionVectorObserved fullCovariateVectorObserved} (Normal prior)}\label{app:betaConditional_normalPrior}
\begin{align}
    p(\betaVector &\vert \lambdaVector, \betaPriorMean, \betaPriorVar, \rewardVectorObserved, \actionVectorObserved, \fullCovariateVectorObserved) \nonumber \\
    \propto& \exp \left\{ -\frac{1}{2} \sum_{\{i: \actionObserved_i = 1\}} \frac{\left( \frac{\rewardObserved_i}{\propensityScore} + \lambda_i -\left( \frac{\rewardObserved_i}{\propensityScore}\right) \actionObserved_i \fullCovariateVectorObserved_i^\top \betaVector \right)^2}{\lambda_i}  \right\} \cdot \exp \left\{ -\frac{1}{2} \sum_{\{i: \actionObserved_i = -1\}} \frac{\left( \frac{\rewardObserved_i}{1-\propensityScore} + \lambda_i -\left( \frac{\rewardObserved_i}{1-\propensityScore}\right) \actionObserved_i \fullCovariateVectorObserved_i^\top \betaVector \right)^2}{\lambda_i}  \right\} \nonumber \\
    &\cdot \exp\left\{-\frac{1}{2} \sum_{j=1}^p \frac{(\beta_j -\mu_{0,j})^2}{\sigma_0^2}\right\} \nonumber \\
    \propto& \exp \left\{ -\frac{1}{2} \sum_{\{i: \actionObserved_i = 1\}} \frac{\left( \lambda_i + \frac{\rewardObserved_i}{\propensityScore}(1 - \actionObserved_i \fullCovariateVectorObserved_i^\top \betaVector) \right)^2}{\lambda_i}  \right\} \cdot \exp \left\{ -\frac{1}{2} \sum_{\{i: \actionObserved_i = -1\}} \frac{\left( \lambda_i + \frac{\rewardObserved_i}{1 - \propensityScore}(1 - \actionObserved_i \fullCovariateVectorObserved_i^\top \betaVector) \right)^2}{\lambda_i}  \right\} \nonumber \\
    &\cdot \exp\left\{-\frac{1}{2} \sum_{j=1}^p \frac{\beta_j^2 -2 \beta_j \mu_{0,j} +\mu_{0,j}^2}{\sigma_0^2}\right\} \nonumber \\
    =& \exp \left\{ -\frac{1}{2} \sum_{\{i: \actionObserved_i = 1\}} \frac{\lambda_i^2 + 2 \lambda_i \frac{\rewardObserved_i}{\propensityScore}(1-\actionObserved_i \fullCovariateVectorObserved_i^\top \betaVector) + \left(\frac{\rewardObserved_i}{\propensityScore} \right)^2 (1 - \actionObserved_i \fullCovariateVectorObserved_i^\top \betaVector)^2) }{\lambda_i}  \right\} \nonumber \\
    & \cdot \exp \left\{ -\frac{1}{2} \sum_{\{i: \actionObserved_i = -1\}} \frac{ \lambda_i^2 + 2 \lambda_i \frac{\rewardObserved_i}{1 - \propensityScore} (1 - \actionObserved_i \fullCovariateVectorObserved_i^\top \betaVector) + \left(\frac{\rewardObserved_i}{1 - \propensityScore} \right)^2(1 - \actionObserved_i \fullCovariateVectorObserved_i^\top \betaVector)^2 }{\lambda_i}  \right\} 
    \exp\left\{-\frac{1}{2} \sum_{j=1}^p \frac{\beta_j^2 -2 \beta_j \mu_{0,j} +\mu_{0,j}^2}{\sigma_0^2}\right\} \nonumber \\
    \propto & \exp \left\{ -\frac{1}{2} \sum_{\{i: \actionObserved_i = 1\}} \left(-2 \frac{\rewardObserved_i}{\propensityScore}\actionObserved_i \fullCovariateVectorObserved_i^\top \betaVector +  \left(\frac{\rewardObserved_i}{\propensityScore} \right)^2 \frac{1}{\lambda_i} (- 2 \actionObserved_i \fullCovariateVectorObserved_i^\top \betaVector + (\actionObserved_i \fullCovariateVectorObserved_i^\top \betaVector)^2)  \right) \right\} \nonumber \\
    & \cdot \exp \left\{ -\frac{1}{2} \sum_{\{i: \actionObserved_i = -1\}} \left(  -2 \frac{\rewardObserved_i}{1 - \propensityScore} \actionObserved_i \fullCovariateVectorObserved_i^\top \betaVector + \left(\frac{\rewardObserved_i}{1 - \propensityScore} \right)^2 \frac{1}{\lambda_i} (- 2\actionObserved_i \fullCovariateVectorObserved_i^\top \betaVector + (\actionObserved_i \fullCovariateVectorObserved_i^\top \betaVector)^2 )  \right) \right\} \nonumber \\
    &\cdot \exp\left\{-\frac{1}{2} \sum_{j=1}^p \frac{\beta_j^2 -2 \beta_j \mu_{0,j}}{\sigma_0^2}\right\} \nonumber \\
    \propto & \exp \left\{ -\frac{1}{2} \sum_{\{i: \actionObserved_i = 1\}} \left(-2 \frac{\rewardObserved_i}{\propensityScore}\actionObserved_i \fullCovariateVectorObserved_i^\top \betaVector \left(1 + \frac{\rewardObserved_i}{\propensityScore \lambda_i} \right) +  \left(\frac{\rewardObserved_i}{\propensityScore} \right)^2 \frac{1}{\lambda_i} (\actionObserved_i \fullCovariateVectorObserved_i^\top \betaVector)^2  \right) \right\} \nonumber \\
    & \cdot \exp \left\{ -\frac{1}{2} \sum_{\{i: \actionObserved_i = -1\}} \left(  -2 \frac{\rewardObserved_i}{1 - \propensityScore} \actionObserved_i \fullCovariateVectorObserved_i^\top \betaVector \left(1 + \frac{1}{(1 - \propensityScore) \lambda_i} \right) + \left(\frac{\rewardObserved_i}{1 - \propensityScore} \right)^2 \frac{1}{\lambda_i}  (\actionObserved_i \fullCovariateVectorObserved_i^\top \betaVector)^2   \right) \right\} \nonumber \\
    &\cdot \exp\left\{-\frac{1}{2} \sum_{j=1}^p \frac{\beta_j^2 -2 \beta_j \mu_{0,j}}{\sigma_0^2}\right\} \nonumber 
\end{align}
Consider the summation inside the exponential of the first term, we have
\begin{align}
    \sum_{\{i: \actionObserved_i = 1\}} & \left(-2 \underbrace{\frac{\rewardObserved_i}{\propensityScore}\actionObserved_i \fullCovariateVectorObserved_i^\top \betaVector \left(1 + \frac{\rewardObserved_i}{\propensityScore \lambda_i} \right)}_{\text{Term 1}} +  \underbrace{\left(\frac{\rewardObserved_i}{\propensityScore} \right)^2 \frac{1}{\lambda_i} (\actionObserved_i \fullCovariateVectorObserved_i^\top \betaVector)^2}_{\text{Term 2}}  \right). \nonumber 
\end{align}
Let $n_1 = \sum_{i = 1}^n \mathbbm{1}(\actionObserved_i = 1)$ Working with the summation over the first term
\begin{align}
    \text{Term 1} =& \sum_{\{\actionObserved_i = 1\}} \frac{\rewardObserved_i}{\propensityScore}\actionObserved_i \fullCovariateVectorObserved_i^\top \betaVector \left(1 + \frac{\rewardObserved_i}{\propensityScore \lambda_i} \right) \nonumber \\
    =& \frac{\rewardObserved_1}{\propensityScore} \actionObserved_1 \fullCovariateVectorObserved_1^\top \betaVector \left(1 + \frac{\rewardObserved_1}{\rho} \lambda_1 \right) + \cdots + \frac{\rewardObserved_{n_1}}{\propensityScore} \actionObserved_{n_1} \fullCovariateVectorObserved_{n_1}^\top \betaVector \left(1 + \frac{\rewardObserved_{n_1}}{\rho} \lambda_{n_1} \right) \nonumber \\
    =& \begin{pmatrix}
        \frac{\rewardObserved_1}{\rho} \left(1 + \frac{\rewardObserved_1}{\rho \lambda_1} \right), \cdots, \frac{\rewardObserved_{n_1}}{\rho} \left(1 + \frac{\rewardObserved_{n_1}}{\rho \lambda_{n_1}} \right) 
    \end{pmatrix}
    \begin{pmatrix}
        \actionObserved_1 \fullCovariateVectorObserved_1^\top \betaVector \\
        \vdots \\
        \actionObserved_{n_1} \fullCovariateVectorObserved_{n_1}^\top \betaVector 
    \end{pmatrix} \nonumber 
\end{align}
Define $\matrixSymbol{X}_1$, $\matrixSymbol{W}_1$, and $\matrixSymbol{R}_1$ as
\begin{align}
    \matrixSymbol{X}_1 &\equiv 
    \begin{pmatrix}
          \actionObserved_1 x_{1,1}    & \cdots    & \actionObserved_1 x_{1,p} \\
         \vdots     &           & \vdots \\
          \actionObserved_{n_1} x_{n_1, 1} & \cdots    & \actionObserved_{n_1} x_{n_1, p}
    \end{pmatrix}_{(n_1 \times p)}, \qquad 
    \matrixSymbol{W}_1 \equiv
    \begin{pmatrix}
        1 + \frac{\rewardObserved_1}{\lambda_1} \\
        \vdots \\
        1 + \frac{\rewardObserved_{n_1}}{\lambda_{n_1}}
    \end{pmatrix}_{(n_1 \times 1)}, \text{ and } \nonumber \\ 
    \matrixSymbol{R}_1 &\equiv diag(r_1/\propensityScore, \ldots, r_{n_1/\propensityScore})_{(n_1 \times n_1)}. \nonumber \\
\end{align}
Then, 
\begin{align}
    \text{Term 1} =&\frac{1}{\propensityScore} \matrixSymbol{W}_1^\top \matrixSymbol{R}_1\matrixSymbol{X}_1 \betaVector
\end{align}
because
\begin{align}
    \matrixSymbol{W}_1^\top \matrixSymbol{R}_1\matrixSymbol{X}_1 \betaVector =&
    \begin{pmatrix}
        1 + \frac{\rewardObserved_1}{\lambda_1} &
        \cdots & 
        1 + \frac{\rewardObserved_{n_1}}{\lambda_{n_1}}
    \end{pmatrix}
    \begin{pmatrix}
        r_1/\propensityScore & \cdots & 0 \\
        \vdots  & \ddots & \vdots \\
        0 & \cdots & r_{n_1}/\propensityScore
    \end{pmatrix} 
    \begin{pmatrix}
          \actionObserved_1 x_{1,1}    & \cdots    & \actionObserved_1 x_{1,p} \\
         \vdots     &           & \vdots \\
          \actionObserved_{n_1} x_{n_1, 1} & \cdots    &  \actionObserved_{n_1} x_{n_1, p}
    \end{pmatrix}
    \begin{pmatrix}
        \beta_1 \\
        \vdots \\
        \beta_p
    \end{pmatrix} \nonumber \\
    =& \frac{1}{\propensityScore} 
    \begin{pmatrix}
        \rewardObserved_1 \left(1 + \frac{\rewardObserved_1}{\lambda_1} \right) 
        & \cdots &
        \rewardObserved_{n_1} \left(1 + \frac{\rewardObserved_{n_1}}{\lambda_{n_1}}\right)
    \end{pmatrix}
    \begin{pmatrix}
        \actionObserved_1 \fullCovariateVectorObserved_1^\top \betaVector \\
        \vdots \\
        \actionObserved_{n_1} \fullCovariateVectorObserved_{n_1}^\top \betaVector
    \end{pmatrix}. \nonumber 
\end{align}

Next, consider the summation over Term 2,
\begin{align}
    \text{Term 2}  =& \sum_{\{i: a_i = -1\}} \left(\frac{\rewardObserved_i}{\propensityScore} \right)^2 \frac{1}{\lambda_i} (\actionObserved_i \fullCovariateVectorObserved_i^\top \betaVector)^2 \nonumber \\
    =& \left(\frac{1}{\propensityScore^2}\right) \left( \frac{\rewardObserved_1^2}{\lambda_1} \right) \actionObserved_1^2 (\fullCovariateVectorObserved_1^\top \betaVector)^2 + \cdots + \left(\frac{1}{\propensityScore^2}\right) \left( \frac{\rewardObserved_{n_1}^2}{\lambda_{n_1}} \right) \actionObserved_{n_1}^2 (\fullCovariateVectorObserved_{n_1}^\top \betaVector)^2. \nonumber 
\end{align}
Observe that 
\begin{align}
    \actionObserved_i^2 &\betaVector^\top \fullCovariateVectorObserved_i \fullCovariateVectorObserved_i^\top \betaVector \nonumber \\ 
    =& \betaVector^\top \begin{pmatrix}
        \actionObserved_i x_{i,1} \\
        \vdots \\
        \actionObserved_i x_{i, p}
    \end{pmatrix}
    \begin{pmatrix}
        \actionObserved_i x_{i,1} &
        \cdots &
        \actionObserved_i x_{i, p}
    \end{pmatrix}
    \betaVector \nonumber \\
    =& \betaVector^\top 
    \begin{pmatrix}
        \actionObserved_i^2 x_{i, 1}^2 & \actionObserved_i^2 x_{i, 1} x_{i, 2} & \cdots & \actionObserved_i^2 x_{i,1} x_{i, p} \\
        \actionObserved_i^2 x_{i, 1} x_{i, 2} & \actionObserved_i^2 x_{i, 2}^2  & \cdots & \actionObserved_i^2 x_{i,2} x_{i, p} \\
        \vdots & \vdots & \ddots & \vdots \\
        \actionObserved_i^2 x_{i, 1} x_{i, p} & \actionObserved_i^2 x_{i, 2} x_{i, p} & \cdots & \actionObserved_i^2 x_{i,p}^2
    \end{pmatrix}
    \betaVector \nonumber  \\
    =& \begin{pmatrix}
        \beta_1 & \cdots & \beta_p
    \end{pmatrix}
    \begin{pmatrix}
        \actionObserved_i^2 x_{i, 1}^2 & \actionObserved_i^2 x_{i, 1} x_{i, 2} & \cdots & \actionObserved_i^2 x_{i,1} x_{i, p} \\
        \actionObserved_i^2 x_{i, 1} x_{i, 2} & \actionObserved_i^2 x_{i, 2}^2  & \cdots & \actionObserved_i^2 x_{i,2} x_{i, p} \\
        \vdots & \vdots & \ddots & \vdots \\
        \actionObserved_i^2 x_{i, 1} x_{i, p} & \actionObserved_i^2 x_{i, 2} x_{i, p} & \cdots & \actionObserved_i^2 x_{i,p}^2
    \end{pmatrix}
    \begin{pmatrix}
        \beta_1 \\
        \vdots \\
        \beta_p
    \end{pmatrix} \nonumber \\
    =& \begin{pmatrix}
        \beta_1 \actionObserved_i^2 x_{i, 1}^2 + \beta_2 \actionObserved_i^2 x_{i, 1} x_{i, 2} + \cdots + \beta_p \actionObserved_i^2 x_{i, 1} x_{i, p}  & \cdots & \beta_1 \actionObserved_i^2 x_{i, 1} x_{i, p} + \beta_2 \actionObserved_i^2 x_{i, 2} x_{i, p} + \cdots + \beta_{p} \actionObserved_i^2 x_{i,p}^2
    \end{pmatrix}
    \begin{pmatrix}
        \beta_1 \\
        \vdots \\
        \beta_p
    \end{pmatrix} \nonumber \\
    =& \beta_1^2 \actionObserved_i^2 x_{i, 1}^2 + \beta_1 \beta_2 \actionObserved_i^2 x_{i, 1} x_{i, 2} + \cdots + \beta_1 \beta_p \actionObserved_i^2 x_{i, 1} x_{i, p} + \cdots + \beta_1 \beta_p \actionObserved_i^2 x_{i, 1} x_{i, p} + \beta_2 \beta_p \actionObserved_i^2 x_{i, 2} x_{i, p} + \cdots + \beta_{p}^2 \actionObserved_i^2 x_{i,p}^2 \nonumber \\
    =& \sum_{j = 1}^p \actionObserved_i^2 x_{i, j}^2 \beta_j^2 + 2\sum_{j = 1}^p \sum_{k \ne j}^p \actionObserved_i^2 x_{i,j}  x_{i, k} \beta_{j} \beta_{k} \nonumber \\
    =& (\actionObserved_i x_{i, 1} \beta_1 + \cdots \actionObserved_i x_{i, p} \beta_p) \cdot \actionObserved_i (x_{i, 1} \beta_1 + \cdots \actionObserved_i x_{i, p} \beta_p) \nonumber \\
     =& (\actionObserved_i \fullCovariateVectorObserved_i^\top \betaVector) \cdot (\actionObserved_i \fullCovariateVectorObserved_i^\top \betaVector) \nonumber \\
     =& (\actionObserved_i \fullCovariateVectorObserved_i^\top \betaVector)^2. \nonumber
\end{align}
Also observe that
\begin{align}
    \betaVector^\top \matrixSymbol{X}_1^\top \matrixSymbol{X}_1 \betaVector =& \betaVector^\top
    \begin{pmatrix}
        \actionObserved_1 x_{1, 1} & \cdots & \actionObserved_{n_1} x_{n_1, 1} \\
        \vdots & \cdots & \vdots \\
        \actionObserved_1 x_{1, p} & \cdots & \actionObserved_{n_1} x_{n_1, p}
    \end{pmatrix}
    \begin{pmatrix}
        \actionObserved_1 x_{1, 1} & \cdots & \actionObserved_1 x_{1, p} \\
        \vdots & \cdots & \vdots \\
        \actionObserved_{n_1} x_{n_1, 1} & \cdots & \actionObserved_{n_1} x_{n_1, p}
    \end{pmatrix}
    \betaVector \nonumber \\
    =& \begin{pmatrix}
        \actionObserved_1 \betaVector^\top \fullCovariateVectorObserved_1 & \cdots & \actionObserved_{n_1} \betaVector^\top \fullCovariateVectorObserved_{n_1}
    \end{pmatrix}
    \begin{pmatrix}
        \actionObserved_1 \fullCovariateVectorObserved_1^\top \betaVector \\
        \cdots \\
        \actionObserved_{n_1} \fullCovariateVectorObserved_{n_1}^\top \betaVector
    \end{pmatrix} \nonumber \\
    =& \begin{pmatrix}
        \actionObserved_1^2 \betaVector^\top \fullCovariateVectorObserved_1 \fullCovariateVectorObserved_1^\top \betaVector \\
        \vdots \\
        \actionObserved_{n_1} \betaVector^\top \fullCovariateVectorObserved_{n_1} \fullCovariateVectorObserved_{n_1}^\top \betaVector.
    \end{pmatrix}
\end{align}
Define $\matrixSymbol{\Lambda}_1 = diag(\lambda_1, \ldots, \lambda_{n_1})$. Then
\begin{align}
    \text{Term 2} =& \left(\frac{1}{\propensityScore^2}\right) \left( \frac{\rewardObserved_1^2}{\lambda_1} \right) \actionObserved_1^2 (\fullCovariateVectorObserved_1^\top \betaVector)^2 + \cdots + \left(\frac{1}{\propensityScore^2}\right) \left( \frac{\rewardObserved_{n_1}^2}{\lambda_{n_1}} \right) \actionObserved_1^2 (\fullCovariateVectorObserved_{n_1}^\top \betaVector)^2 \nonumber \\
    =& \left(\frac{1}{\propensityScore^2} \right) \begin{pmatrix}
         \left(\frac{\rewardObserved_1^2}{\lambda_1}\right)  & \cdots & \left(\frac{\rewardObserved_{n_1}^2}{\lambda_{n_1}}\right) 
    \end{pmatrix}
    \begin{pmatrix}
        \actionObserved_1^2 \betaVector^\top \fullCovariateVectorObserved_1 \fullCovariateVectorObserved_1^\top \betaVector \\
        \vdots \\
        \actionObserved_{n_1}^2 \betaVector^\top \fullCovariateVectorObserved_{n_1} \fullCovariateVectorObserved_{n_1}^\top \betaVector
    \end{pmatrix} \nonumber \\
    =& \left(\frac{1}{\propensityScore^2} \right) \begin{pmatrix}
         \left(\frac{\rewardObserved_1^2}{\lambda_1}\right)  & \cdots & \left(\frac{\rewardObserved_{n_1}^2}{\lambda_{n_1}}\right) 
    \end{pmatrix} \betaVector^\top \matrixSymbol{X}_1^\top \matrixSymbol{X}_1 \betaVector \nonumber \\
    =& \betaVector^\top \matrixSymbol{X}_1^\top \matrixSymbol{R}_1^\top \matrixSymbol{\Lambda}_1^{-1} \matrixSymbol{R}_1 \matrixSymbol{X}_1 \betaVector.
\end{align}

Let $n_{-1} = \sum_{i = 1}^n \mathbbm{1}(\actionObserved_i = -1)$. Similarly, define $\matrixSymbol{X}_{-1}$, $\matrixSymbol{W}_{-1}$, $\matrixSymbol{R}_{-1}$, and $\boldsymbol{\Lambda}_{-1}$ as
\begin{align}
    \matrixSymbol{X}_{-1} &\equiv 
    \begin{pmatrix}
         a_1 x_{1,1}    & \cdots    & a_1 x_{1,p} \\
         \vdots     &           & \vdots \\
         a_{n_{-1}} x_{n_{-1}, 1} & \cdots    & a_{n_{-1}} x_{n_{-1}, p}
    \end{pmatrix}_{(n_{-1} \times p)}, \qquad 
    \matrixSymbol{W}_{-1} \equiv
    \begin{pmatrix}
        1 + \frac{\rewardObserved_1}{\lambda_1} \\
        \vdots \\
        1 + \frac{\rewardObserved_{n_{-1}}}{\lambda_{n_{-1}}}
    \end{pmatrix}_{(n_{-1} \times 1)},  \nonumber \\ 
    \matrixSymbol{R}_{-1} &\equiv diag(r_1/(1-\propensityScore), \ldots, r_{n_{-1}}/(1-\propensityScore))_{(n_{-1} \times n_{-1})}, \text{ and }\quad \matrixSymbol{\Lambda}_{-1} = diag(\lambda_1, \ldots, \lambda_{n_{-1}}). \nonumber 
\end{align}
Additionally define $\matrixSymbol{\Sigma} \equiv diag(\sigma_1, \ldots, \sigma_p) $
so that we can write
\begin{align}
     \exp\left\{-\frac{1}{2} \sum_{j=1}^p \frac{\beta_j^2 -2 \beta_j \mu_{0,j}}{\sigma_0^2}\right\} =& \exp \left\{-\frac{1}{2} (\betaVector^\top \matrixSymbol{\Sigma}^{-1}\betaVector -2 \betaPriorMean^\top \matrixSymbol{\Sigma}^{-1}\betaVector) \right\}.
\end{align}
Thus, we have that
\begin{align}
   p(\betaVector &\vert \lambdaVector, \omegaVector, \rewardVectorObserved, \actionVectorObserved, \fullCovariateVectorObserved) \nonumber \\
    \propto& \exp\left\{-\frac{1}{2}  \left(-2  \matrixSymbol{W}_1^\top \matrixSymbol{R}_1\matrixSymbol{X}_1 \betaVector + \betaVector^\top \matrixSymbol{X}_1^\top \matrixSymbol{R}_1^\top \matrixSymbol{\Lambda}_1^{-1} \matrixSymbol{R}_1 \matrixSymbol{X}_1 \betaVector \right) \right\} \nonumber \\
    &\cdot \exp \left\{-\frac{1}{2} \left(-2  \matrixSymbol{W}_{-1}^\top \matrixSymbol{R}_{-1} \matrixSymbol{X}_{-1} \betaVector +  
    \betaVector^\top \matrixSymbol{X}_{-1}^\top \matrixSymbol{R}_{-1}^\top \matrixSymbol{\Lambda}_{-1}^{-1} \matrixSymbol{R}_{-1} \matrixSymbol{X}_{-1} \betaVector 
    \right) \right\} \nonumber \\
    &\cdot \exp \left\{ -\frac{1}{2} (-2 \betaPriorMean^\top \matrixSymbol{\Sigma}^{-1} \betaVector + \betaVector^\top  \matrixSymbol{\Sigma}^{-1} \betaVector) \right\} \nonumber \\
    =& \exp \bigg\{-\frac{1}{2} \Bigg[\betaVector^\top \underbrace{\left(\matrixSymbol{X}_1^\top \matrixSymbol{R}_1^\top \matrixSymbol{\Lambda}_1^{-1} \matrixSymbol{R}_1 \matrixSymbol{X}_1 + \matrixSymbol{X}_{-1}^\top \matrixSymbol{R}_{-1}^\top \matrixSymbol{\Lambda}_{-1}^{-1} \matrixSymbol{R}_{-1} \matrixSymbol{X}_{-1} +  \matrixSymbol{\Sigma}^{-1}\right)}_{\equiv B_1^{-1}}  \betaVector \nonumber \\
    & -2 (\underbrace{\matrixSymbol{W}_1^\top \matrixSymbol{R}_1\matrixSymbol{X}_1 + \matrixSymbol{W}_{-1}^\top \matrixSymbol{R}_{-1} \matrixSymbol{X}_{-1} + \betaPriorMean^\top \matrixSymbol{\Sigma}^{-1} }_{\equiv b_1})\betaVector \Bigg] \bigg\} \nonumber \\
    =& \exp \left\{-\frac{1}{2} (\betaVector - B_1 b_1)^\top B^{-1} (\betaVector - B_1 b_1) - b_1^\top B_1 b_1 \right\} \nonumber \\
    \propto& \exp \left\{-\frac{1}{2} (\betaVector - B_1 b_1)^\top B_1^{-1} (\betaVector - B_1 b_1) \right\}. \nonumber
\end{align}
The conditional distribution of $\betaVector$ given $\lambdaVector$ is multivariate normal with mean $B_1 b_1$ and variance-covariance matrix $B_1$.

\subsubsection{Conditional distribution of \texorpdfstring{$\betaVector \vert \lambdaVector, \omegaVector, \rewardVectorObserved, \actionVectorObserved, \fullCovariateVectorObserved$}{betaVector given lambdaVector omegaVector rewardVectorObserved actionVectorObserved fullCovariateVectorObserved} (Exponential power prior)}
\begin{align}
    p(\betaVector &\vert \lambdaVector, \omegaVector, \rewardVectorObserved, \actionVectorObserved, \fullCovariateVectorObserved) \nonumber \\
    \propto& \exp \left\{ -\frac{1}{2} \sum_{\{i: \actionObserved_i = 1\}} \frac{\left( \frac{\rewardObserved_i}{\propensityScore} + \lambda_i -\left( \frac{\rewardObserved_i}{\propensityScore}\right) \actionObserved_i \fullCovariateVectorObserved_i^\top \betaVector \right)^2}{\lambda_i}  \right\} \cdot \exp \left\{ -\frac{1}{2} \sum_{\{i: \actionObserved_i = -1\}} \frac{\left( \frac{\rewardObserved_i}{1-\propensityScore} + \lambda_i -\left( \frac{\rewardObserved_i}{1-\propensityScore}\right) \actionObserved_i \fullCovariateVectorObserved_i^\top \betaVector \right)^2}{\lambda_i}  \right\} \nonumber \\
    &\cdot \exp\left\{-\frac{1}{2\nu^2} \sum_{j = 1}^p \frac{\beta_j^2}{\sigma_j^2 \omega_j} \right\} \nonumber \\
    \propto & \exp \left\{ -\frac{1}{2} \sum_{\{i: \actionObserved_i = 1\}} \left(-2 \frac{\rewardObserved_i}{\propensityScore}\actionObserved_i \fullCovariateVectorObserved_i^\top \betaVector \left(1 + \frac{\rewardObserved_i}{\propensityScore \lambda_i} \right) +  \left(\frac{\rewardObserved_i}{\propensityScore} \right)^2 \frac{1}{\lambda_i} (\actionObserved_i \fullCovariateVectorObserved_i^\top \betaVector)^2  \right) \right\} \nonumber \\
    & \cdot \exp \left\{ -\frac{1}{2} \sum_{\{i: \actionObserved_i = -1\}} \left(  -2 \frac{\rewardObserved_i}{1 - \propensityScore} \actionObserved_i \fullCovariateVectorObserved_i^\top \betaVector \left(1 + \frac{1}{(1 - \propensityScore) \lambda_i} \right) + \left(\frac{\rewardObserved_i}{1 - \propensityScore} \right)^2 \frac{1}{\lambda_i}  (\actionObserved_i \fullCovariateVectorObserved_i^\top \betaVector)^2   \right) \right\} \nonumber \\
    & \cdot \exp\left\{-\frac{1}{2 \nu^2} \sum_{j = 1}^p \frac{\beta_j^2}{\sigma_j^2 \omega_j} \right\} \nonumber
\end{align}
The summation inside the first and second exponential terms are the same as in the derivation under the normal distribution prior for $\betaVector$ (\cref{app:betaConditional_normalPrior}). Letting $\matrixSymbol{\Omega} \equiv diag(\omega_1, \ldots, \omega_p)_{(p \times p)}$, we can write
\begin{align}
     \sum_{j = 1}^p \frac{\beta_j^2}{\sigma_j^2 \omega_j} 
     =&
     \begin{pmatrix}
         1/(\sigma_1^2 \omega_1) &\cdots & 1/(\sigma_p^2 \omega_p)
     \end{pmatrix} 
     \begin{pmatrix}
         \beta_1^2 \\ \vdots \\ \beta_p^2
     \end{pmatrix} \nonumber \\
     =&  \begin{pmatrix}
         \beta_1^2 & \cdots & \beta_p^2
     \end{pmatrix}
     \begin{pmatrix}
         1/(\sigma_1^2 \omega_1) &\cdots &0 \\
         \vdots & \ddots & \vdots \\
         0 & \cdots & 1/(\sigma_p^2 \omega_p) 
     \end{pmatrix}
     \begin{pmatrix}
         \beta_1 \\ \vdots \\ \beta_p
     \end{pmatrix} \nonumber \\
     =& 
     \begin{pmatrix}
         \beta_1^2 & \cdots & \beta_p^2
     \end{pmatrix}
     \begin{pmatrix}
         1/\sigma_1^2 &\cdots &0 \\
         \vdots & \ddots & \vdots \\
         0 & \cdots & 1/\sigma_p^2
     \end{pmatrix}
     \begin{pmatrix}
         1/\omega_1 & \cdots & 0 \\
         \vdots & \ddots & \vdots \\
         0 & \cdots & 1/\omega_p
     \end{pmatrix}
     \begin{pmatrix}
         \beta_1 \\ \vdots \\ \beta_p
     \end{pmatrix} \nonumber \\
     =& 
     \betaVector^\top \matrixSymbol{\Omega}^{-1} \matrixSymbol{\Sigma}^{-1} \betaVector \nonumber 
\end{align}
Thus, we have that
\begin{align}
   p(\betaVector &\vert \lambdaVector, \omegaVector, \rewardVectorObserved, \actionVectorObserved, \fullCovariateVectorObserved) \nonumber \\
    \propto& \exp\left\{-\frac{1}{2}  \left(-2  \matrixSymbol{W}_1^\top \matrixSymbol{R}_1\matrixSymbol{X}_1 \betaVector + \betaVector^\top \matrixSymbol{X}_1^\top \matrixSymbol{R}_1^\top \matrixSymbol{\Lambda}_1^{-1} \matrixSymbol{R}_1 \matrixSymbol{X}_1 \betaVector \right) \right\} \nonumber \\
    &\cdot \exp \left\{-\frac{1}{2} \left(-2  \matrixSymbol{W}_{-1}^\top \matrixSymbol{R}_{-1} \matrixSymbol{X}_{-1} \betaVector +  
    \betaVector^\top \matrixSymbol{X}_{-1}^\top \matrixSymbol{R}_{-1}^\top \matrixSymbol{\Lambda}_{-1}^{-1} \matrixSymbol{R}_{-1} \matrixSymbol{X}_{-1} \betaVector 
    \right) \right\} \nonumber \\
    &\cdot \exp \left\{ -\frac{1}{2} \nu^{-2} \betaVector^\top \matrixSymbol{\Omega}^{-1} \matrixSymbol{\Sigma}^{-1} \betaVector \right\} \nonumber \\
    =& \exp \bigg\{-\frac{1}{2} \Bigg[\betaVector^\top \underbrace{\left(\matrixSymbol{X}_1^\top \matrixSymbol{R}_1^\top \matrixSymbol{\Lambda}_1^{-1} \matrixSymbol{R}_1 \matrixSymbol{X}_1 + \matrixSymbol{X}_{-1}^\top \matrixSymbol{R}_{-1}^\top \matrixSymbol{\Lambda}_{-1}^{-1} \matrixSymbol{R}_{-1} \matrixSymbol{X}_{-1} + \nu^{-2} \matrixSymbol{\Omega}^{-1} \matrixSymbol{\Sigma}^{-1}\right)}_{\equiv B_2^{-1}}  \betaVector \nonumber \\
    & -2 (\underbrace{\matrixSymbol{W}_1^\top \matrixSymbol{R}_1\matrixSymbol{X}_1 + \matrixSymbol{W}_{-1}^\top \matrixSymbol{R}_{-1} \matrixSymbol{X}_{-1} }_{\equiv b_2})\betaVector \Bigg] \bigg\} \nonumber \\
    =& \exp \left\{-\frac{1}{2} (\betaVector - B_2 b_2)^\top B_2^{-1} (\betaVector - B_2 b_2) - b_2^\top B_2 b_2 \right\} \nonumber \\
    \propto& \exp \left\{-\frac{1}{2} (\betaVector - B_2 b_2)^\top B_2^{-1} (\betaVector - B_2 b_2) \right\}. \nonumber
\end{align}
The conditional distribution of $\betaVector$ given $\lambdaVector$, $\omegaVector$, and is multivariate normal with mean $B_2 b_2$ and variance-covariance matrix $B_2$.

\end{document}